# Molecular Dynamics simulations of Doxorubicin in sphingomyelin-based lipid membranes

Paulo Siani,[1] Edoardo Donadoni,[1] Lorenzo Ferraro,[1] Francesca Re,[2,3] Cristiana Di Valentin[1,3]

[1] Department of Materials Science, University of Milano-Bicocca
via R. Cozzi 55, 20125 Milano Italy

[2] School of Medicine and Surgery, University of Milano-Bicocca,
via Raoul Follereau 3, Vedano al Lambro, MB 20854, Italy

[3] BioNanoMedicine Center NANOMIB, University of Milano-Bicocca

**Abstract**

Doxorubicin (DOX) is one of the most efficient antitumor drugs employed in numerous cancer therapies. Its incorporation into lipid-based nanocarriers, such as liposomes, improves the drug targeting into tumor cells and reduces drug side effects. The carriers' lipid composition is expected to affect the interactions of DOX and its partitioning into liposomal membranes. To get a rational insight into this aspect and determine promising lipid compositions, we use numerical simulations, which provide unique information on DOX-membrane interactions at the atomic level of resolution. In particular, we combine classical molecular dynamics simulations and free energy calculations to elucidate the mechanism of penetration of a protonated Doxorubicin molecule ($DOX^+$) into potential liposome membranes, here modeled as lipid bilayers based on mixtures of phosphatidylcholine (PC), sphingomyelin (SM) and cholesterol lipid molecules, of different compositions and lipid phases. Moreover, we analyze $DOX^+$ partitioning into relevant regions of SM-based lipid bilayer systems using a combination of free energy methods. Our results show that $DOX^+$ penetration and partitioning are facilitated into less tightly packed SM-based membranes and are dependent on lipid composition. This work paves the way to further investigations of optimal formulations for lipid-based carriers, such as those associated with pH-responsive membranes.

1. **Introduction**

Doxorubicin (DOX) is a chemotherapy drug belonging to the family of the anthracyclines, some of the most prescribed and efficient anticancer drugs in treating breast carcinoma, osteosarcoma, acute leukemia, brain tumors, among others.[1] The current and most accepted hypothesis of the DOX cytotoxic mechanism suggests a first step of drug intercalation within the DNA double helix leading to its damage, and consequently, inhibiting further tumoral cell growing and replication.[2] However, the cytotoxicity of DOX is nonspecific and normal cells, such as those in the heart tissue [3-5], are also affected. One way to overcome this limitation is to design lipid-based carriers to improve the targeting and efficacy of DOX, thus reducing its cardiotoxicity.

Liposomes, as drug carriers, have phospholipids and cholesterol as the most fundamental structural components in their formulations.[6][7] More recently, the PEGylation of these liposomes, the so-called *stealth* liposomes, has been a significant step forward towards a more specific targeting on tumoral issues.[8-10] For instance, the FDA-approved and clinically widely applied DOXIL® seems to favor a specific releasing of DOX into the interstitial fluid of tumor cells instead of the normal cells, with further uptake of the drug through passive diffusion, although the exact mechanism is still controversial.[11]

In general, liposomes have been used to improve the therapeutic index of new or established drugs by modifying drug absorption, reducing metabolism, prolonging biological half-life or reducing toxicity. Drug distribution is then primarily controlled by the carrier properties and no longer by the physico-chemical characteristics of the drug substance only. Conventional or polymer-grafted liposomal formulations have been usually based on vesicles composed of phosphatidyl-phospholipids mixtures that, however, as one of their drawbacks, are susceptible to acid- or enzymatic-catalyzed hydrolysis of phosphatidylcholines, leading to liposome degradation and drug leakage.[12-15] Sphingomyelins (SM)-based liposomes, known as *sphingosomes*, have amide-linked fatty acids rather than ester-linked ones, turning them out to be less susceptible to hydrolysis, preventing their chemical degradation and therefore improving the drug circulation lifetime and the liposomes' *in vitro* stability.[12,13,16]

Incorporating antitumoral drugs such as DOX into liposome vesicles is mainly driven by drug-liposome interactions. The optimal balance in these molecular interactions plays an essential role in improving the pharmacokinetics and the antitumoral activity in any successful liposomal carrier formulation. Antitumoral drugs, depending on their hydrophobic/hydrophilic characteristics, are located in the aqueous core, inside the lipid bilayers of liposomes or at the bilayer interface and these characteristics directly affect the release profile of the entrapped tumor-targeting drug.[17] Hence, a molecular understanding of the drug-membrane interactions and the drug partitioning into lipid bilayer models, whose lipid compositions are shared by both liposome carriers and cell membranes, is of the utmost importance. Numerical molecular dynamics (MD) simulations offer useful tools to accomplish this aim, providing relevant information at an atomistic resolution level, which can hardly be achieved using experimental techniques.

Due to previous efforts to parametrize and validate atomistic force-field (FF) for SM and phosphatidylcholine (PC) lipid bilayer models[18-29], nowadays, state-of-the-art FF-based MD simulations can reap the fruits of these fundamental works and better understand the role of anthracyclines drugs in these membranes.[30] Yacoub et al.[31] investigated DOX penetration in mixed dipalmitoylphosphatidylcholine (DPPC)/cholesterol lipid bilayers at different cholesterol contents through unrestrained and constrained MD simulations. They found out that the free energy barrier of DOX translocation across the DPPC bilayers increases as cholesterol content increases. Another important finding in their work is the facilitation of water penetration into lipid bilayers, mainly driven by the polar moieties of DOX. Reis et al.[32] investigated, in a combined experimental and MD simulation study, the interaction of DOX with liposomes consisting of mixtures of dimyristoylphosphocholine (DMPC)/SM and DMPC/SM/cholesterol lipids. They noticed that DOX shows similar partitioning in both DMPC/SM and DMPC/SM/cholesterol lipid bilayers and a preferential DOX location in these bilayers' headgroups region. However, so far, no in-depth MD studies investigating the DOX interaction with pure and cholesterol-enriched SM bilayers, which represent the standard lipid composition in many sphingosomes formulations, have been carried out. To better understand the DOX interactions with PC and SM-based lipid membranes, we carried out classical MD simulations of hydrated DPPC and N-palmitoylsphingomyelin (PSM) lipid bilayer models in the presence of this drug.

This paper clarifies the main differences of DOX interaction in both pure and cholesterol-enriched PSM and the DPPC lipid bilayers, through an extensive analysis of structural, dynamical and thermodynamic properties, which are estimated by means of force-field MD simulations. We first present a systematic comparison of the penetration and of the location of a protonated Doxorubicin molecule (hereafter DOX$^+$, Scheme 1) in distinct lipid phases of DPPC and PSM bilayers in either the presence or in the absence of cholesterol. Secondly, through free energy calculation methods, we investigate how the presence of cholesterol affects the DOX$^+$ partitioning in the solid- and liquid-ordered lipid phases of both pure PSM and PSM/cholesterol bilayers, respectively. Finally, we discuss our results in light of available experimental and theoretical findings.

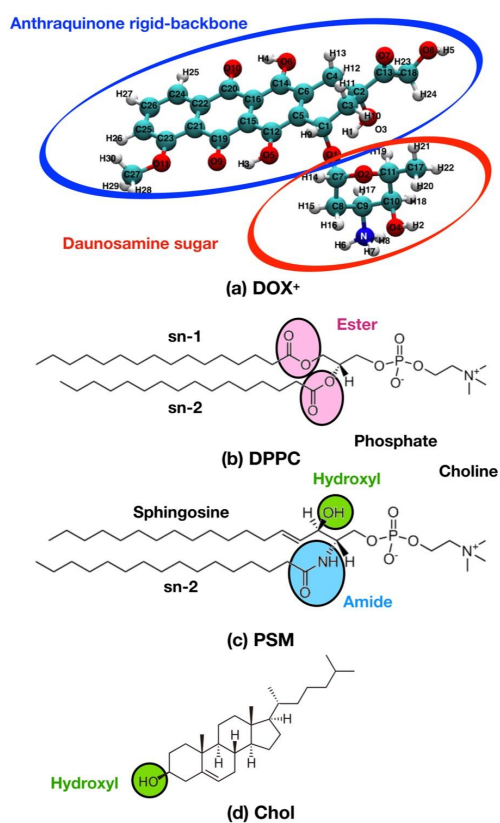

Scheme 1. Chemical structures of (a) protonated Doxorubicin molecule (DOX$^+$) with the atom-type labels, (b) dipalmitoylphosphatidylcholine (DPPC) molecule, (c) N-palmitoylsphingomyelin (PSM) molecule and (d) cholesterol (Chol) molecule.

Relevant functional groups of the lipid molecules are written and differences between DPPC, PSM and Chol molecules are highlighted with colors.

## 2. Computational details

DOX$^+$ was studied in two simulation sets. The first simulation set comprised a pure lipid bilayer containing 128 PSM lipids with 7040 water molecules simulated in either the solid-ordered (T = 303.15 K) or the liquid-disordered (T = 323.15 K) phases, and a second lipid bilayer composed of a binary mixture of 64 PSM lipids and 64 cholesterol molecules hydrated with 7040 water molecules simulated in the liquid-ordered lipid phase (T = 303.15 K). The second simulation set comprised a pure DPPC bilayer containing 128 lipids with 5310 water molecules simulated in the solid-ordered (T = 303.15 K) and the liquid-disordered (T = 323.15 K) lipid phases, and a second lipid bilayer system composed of a binary mixture containing 64 DPPC lipids and 64 cholesterol molecules hydrated with 4476 water molecules, simulated in the liquid-ordered lipid phase (T = 303.15 K). A single DOX$^+$ molecule was added in the bulk-water phase of all the above-mentioned systems, far from the lipid bilayer center at a minimum distance of 35 Å, where water molecules either overlapping or lying at distances shorter than 2.0 Å from the solute were removed. The initial structure of DOX$^+$ was obtained on the web-server MolView v.2.4 (molview.org).[33] Partial atomic charges were fitted according to the CHARMM protocol,[34] in order to accurately describe the intermolecular interactions between DOX$^+$ and TIP3P water molecules at the HF/6-31G* level of theory. To maintain the overall electroneutrality of the systems, one chloride counterion was randomly added in the bulk-water phase to counter-balance the positive net charge of DOX$^+$. All MD simulations used the CGenFF[35-37] for DOX$^+$, CHARMM36 FF for lipids[25,38,39] and CHARMM TIP3P water model[40-42]. Starting-point structures for the lipid bilayer systems were built using the CHARMM-GUI *Membrane Builder*[43-47], and then the standard *Membrane Builder* minimization and equilibration steps were carried out to remove the restraints gradually. If not mentioned otherwise, MD simulations were carried out with pressure (P = 1 atm) and temperature (T = 303.15 K) held constant by using a Langevin friction force with the damping coefficient at 0.1 ps$^{-1}$ and a semi-isotropic Nosé-Hoover Langevin-piston scheme[48,49] with a piston period of 50 fs and a piston decay of 25 fs, respectively. A 2 fs time step was used to integrate the equations of motion by means of the impulse-based Verlet-I/r-RESPA integrator[50] in an isothermal-isobaric ensemble (NPT) under periodic boundary conditions. The Particle Mesh Ewald (PME) scheme[51] was used to evaluate the long-range electrostatic interactions using a real-space cutoff of 12 Å. Lennard-Jones interactions were truncated with a 12 Å cutoff with a switching function applied beyond 10 Å. A combination of SHAKE/RATTLE[52,53] and SETTLE algorithms[54] were used to impose holonomic constraints on all covalent bonds including hydrogen atoms. All MD simulations were carried out using the NAMD2.13 program[55] and its CUDA module for GPU-accelerated calculations.

## 3. Results

In this work, we study the interactions of $DOX^+$ and lipid membranes composed of DPPC, DPPC/cholesterol, PSM, and PSM/cholesterol by means of force-field based MD simulations. In the following, we will first present a systematic comparison of $DOX^+$ penetration and location in distinct lipid phases of DPPC (Section 3.1) and PSM (Section 3.2) bilayers, both in the absence and in the presence of cholesterol. Next (Section 3.2.3), we will analyze, based on free energy calculations, how the presence of cholesterol affects the $DOX^+$ partitioning in solid-ordered PSM ($PSM^{So}$) and in liquid-ordered PSM/cholesterol ($PSM^{Lo}$/cholesterol) bilayers. Finally, in the following Section 4, we will discuss our results in the light of current literature.

### 3.1. DPPC bilayers

#### 3.1.1. Doxorubicin in pure DPPC bilayers

In this section, we investigate the interaction of a single $DOX^+$ molecule and pure DPPC bilayers in the solid-ordered ($DPPC^{So}$) and the liquid-disordered ($DPPC^{Ld}$) lipid phases. To this end, we examine energetic, structural, and dynamical parameters of $DOX^+$ interacting with and penetrating these lipid bilayers. To obtain the location of $DOX^+$ in both So (T = 303.15 K) and Ld (T = 323.15 K) lipid phases in pure DPPC bilayers, we compute the electron density profile along the z-direction to the bilayer normal (For further details, please see Section 3.3, Supplementary Material). Figure 1 shows the electron density profile of DOX$^+$ and the main lipid groups averaged over the last 100 ns of a total MD production time of 500 ns.

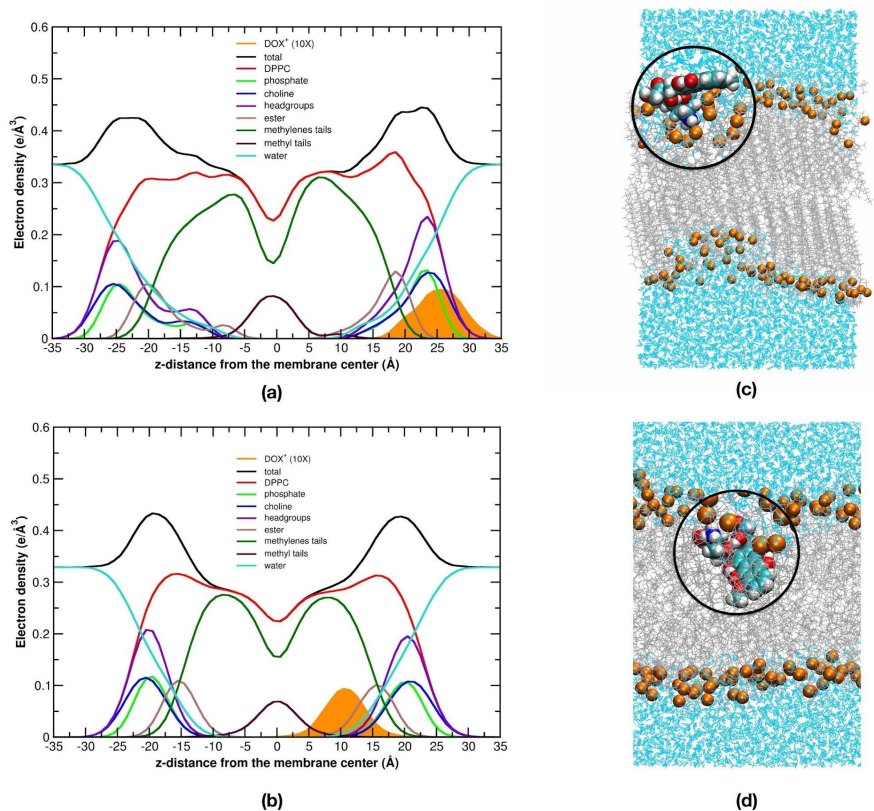

Figure 1. (a,b) Electron density profiles of the main lipid groups in (a) DPPC$^{So}$/DOX$^+$ and (b) DPPC$^{Ld}$/DOX$^+$ systems over the last 100 ns MD production phase. (c,d) Final snapshot from the MD production phase of (c) DPPC$^{So}$/DOX$^+$ and (d) DPPC$^{Ld}$/DOX$^+$. Water molecules are shown in blue, DPPC lipids in grey and DPPC phosphate groups in orange. For the sake of clarity, the intensity of the DOX$^+$ electron density profile is enhanced by 10 times.

In Figure 1a, we observe DOX$^+$ at about 25.5 Å from the DPPC$^{So}$ bilayer center, whereas this distance lowers to 10.5 Å when considering the DPPC$^{Ld}$ bilayer. This result shows that DOX$^+$ has enhanced penetration in the DPPC$^{Ld}$ bilayer, evidenced by the symmetric peak at the interfacial region between the apolar hydrocarbon chains and the polar headgroups in Figure 1b. The visual inspection of the last MD snapshot reveals that the DOX$^+$ rigid backbone is completely dragged towards the inner regions of the DPPC$^{Ld}$ bilayer (Figure 1d), whereas it remains at outer regions in the DPPC$^{So}$ bilayer (Figure 1c). In both cases, the DOX-NH$_3^+$ group maintains a strong interaction with the polar groups in DPPC bilayers.

To quantify the driving force in drug-membrane interactions, we compute the non-bonded interactions for the DOX$^+$/DPPC$^{So}$ and DOX$^+$/DPPC$^{Ld}$ pairs, breaking down the total interaction energy into their electrostatic and vdW contributions over the 500 ns of MD production phase. Figure 2 shows the electrostatic and vdW interactions of DOX$^+$ with the DPPC$^{So}$ (Figure 2a) and DPPC$^{Ld}$ (Figure 2b) bilayers. To track the location of DOX$^+$, we also include the z-distance of the center of mass (C.O.M.) of DOX$^+$ from the bilayer center over the entire production phases. (Figures 2c and 2d).

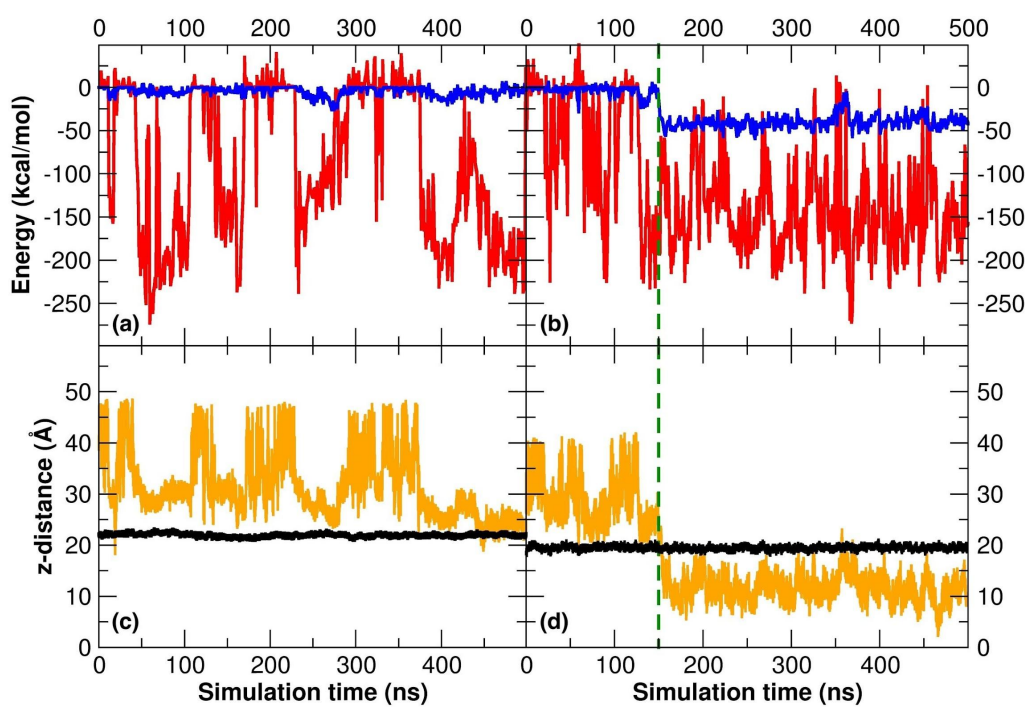

Figure 2. (a,b) Electrostatic (red) and vdW (blue) interaction energies for the (a) DOX$^+$/DPPC$^{So}$ and (b) DOX$^+$/DPPC$^{Ld}$ pairs. (c,d) z-component of the C.O.M distance of DOX$^+$ (orange) and DPPC-PO$_4^-$ (black) groups from the bilayer center in (c) DOX$^+$/DPPC$^{So}$ and (d) DOX$^+$/DPPC$^{Ld}$ systems over 500 ns of MD production simulation.

The interactions between the DOX$^+$/DPPC$^{So}$ and DOX$^+$/DPPC$^{Ld}$ pairs, as revealed in Figure 2, are mainly of electrostatic nature. This fact is intimately correlated with the molecular closeness between the highly polar groups of DOX-NH$_3^+$ and DPPC-PO$_4^-$, as previously seen in Figure 1.

In the DOX$^+$/DPPC$^{Ld}$ case, the hydrophobic contribution to the total interaction energy is negligible up to 150 ns, at which point a stabilization due to the vdW forces starts showing up, due to the non-polar interaction between the DOX$^+$ rigid backbone and the DPPC hydrocarbon tails. We identify no similar hydrophobic stabilization in the DOX$^+$/DPPC$^{So}$ system, where the main contribution to the total energy comes from electrostatic interactions. The above is in line with estimations of the z-distance C.O.M of DOX$^+$ from the bilayer center (Figures 2c and 2d). We see a drop in the z-distance C.O.M of DOX$^+$ towards the DPPC$^{Ld}$ bilayer core at about 150 ns of MD simulation (Figure 2d), whereas no such an event is observed during the DOX$^+$/DPPC$^{So}$ simulation (Figure 2c).

Both electrostatic and vdW energies are intimately related to the H-bond formations between DOX$^+$ and the main-polar groups in lipid bilayers. Hence, to obtain further insights into the molecular nature of drug-membrane interactions, we now analyze the RDF and the H-bond formation between the DOX-NH$_3^+$ group and phosphate (from now on DPPC-PO$_4^-$) and ester (from now on DPPC-ROR$^j$) groups of DPPC lipids, together with the drug/water interactions.

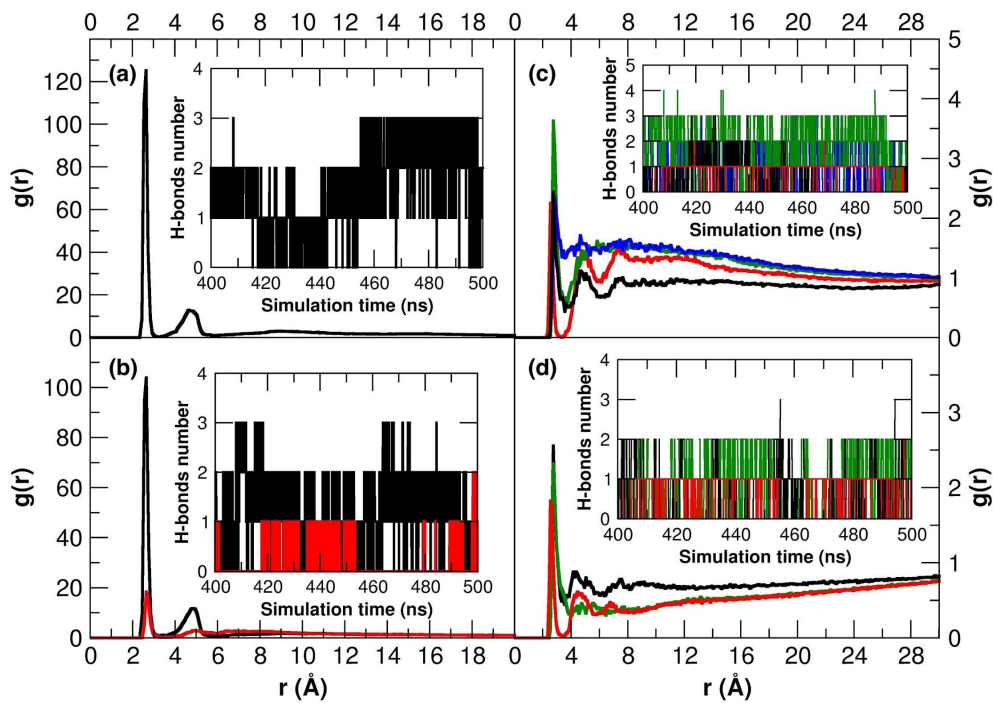

Figure 3. (a,b) Radial distribution function and number of H-bond formation (inset) for the DOX-$NH_3^+$/DPPC-$PO_4^-$ (black) and DOX-$NH_3^+$/ DPPC-ROR$^|$ (red) pairs in (a) DOX$^+$/DPPC$^{So}$ and (b) DOX$^+$/DPPC$^{Ld}$ systems over the last 100 ns MD production phase. (c,d) Radial distribution function and number of H-bond formation (inset) for the DOX-$NH_3^+$/water (black), DOX-O1H/water (red), DOX-O3H/water (green) and DOX-O8H/water (blue) pairs in (c) DOX$^+$/DPPC$^{So}$ and (d) DOX$^+$/DPPC$^{Ld}$ systems over the last 100 ns MD production simulation.

Figure 3 reveals that, in both systems, the DOX-$NH_3^+$ group is mainly surrounded by DPPC-$PO_4^-$ groups at an intermolecular distance of about 2.65 Å, which is maintained by the recurrent formation of H-bonds between these two groups over the MD production phase (Figure 3a). In the DOX$^+$/DPPC$^{Ld}$ system, however, the DOX-$NH_3^+$ group also has considerable interaction with the DPPC-ROR$^|$ groups, as supported by Figure 3b.

DOX$^+$ has several polar groups that can establish hydrogen bonds with water molecules, as indicated in the RDF and H-bonds plots of Figure 3c and Figure 3d. In particular, the nitrogen atom of the DOX$^+$ sugar and the O1, O3, O4 and O8 atoms are the most prone to interact with water (please see Scheme 1a for atoms labeling). Moreover, the number of hydrogen bonds is superior in the DPPC$^{So}$ case than in the DPPC$^{Ld}$ one, which is supported by the fact that in the DPPC$^{Ld}$ case, DOX$^+$ spends the last 100 ns of the simulation in a region of the bilayer close to the lipid tails, where the degree of water penetration is low. In the DPPC$^{So}$ case, instead, the drug sets at the interface between the lipid headgroups and the aqueous phase, where it can be more solvated.

In light of the results obtained in this section, we can summarize the main aspects of DOX$^+$ dynamics during the MD simulation as below. In the DPPC$^{So}$/DOX$^+$ system, we observe that DOX$^+$ starts to diffuse within the DPPC$^{So}$ headgroups region at about 400 ns of MD production phase. Then DOX$^+$ is dragged, mainly driven by long-range electrostatic interactions, from the bulk water towards the negatively charged DPPC-PO$_4^-$ headgroups. Along the last 100 ns of MD production phase, DOX$^+$ remains positioned at the membrane-water interface, with its DOX-NH$_3^+$ group pointing towards the DPPC-PO$_4^-$ headgroups, while its rigid backbone resides outer and parallel to the DPPC$^{So}$ bilayer normal. In contrast, we find that DOX$^+$ shows higher penetration through the DPPC$^{Ld}$, as compared to the DPPC$^{So}$ bilayer. DOX$^+$ penetration into the DPPC$^{Ld}$ bilayer follows a three-step process: first, DOX$^+$ diffuses towards the DPPC-PO$_4^-$ headgroups, resembling that observed in the DOX$^+$/DPPC$^{So}$ system, within the simulation time of 100 ns. When we extend the simulation time to 500 ns, we notice the DOX$^+$ rigid backbone reversing its position along the bilayer normal, mainly driven by hydrophobic interactions, and then settling itself, almost parallel, at the nonpolar hydrocarbon region in the DPPC$^{Ld}$ bilayer. This latter event does not occur in the DOX$^+$/DPPC$^{So}$ system, where DOX$^+$ holds its position at the outer region of the bilayer.

### 3.1.2. Doxorubicin in cholesterol-enriched DPPC bilayer

In this section, we investigate the influence of cholesterol on the DOX$^+$ interactions with a binary mixture of DPPC enriched with 50% cholesterol content. To this purpose, we simulate a single DOX$^+$ molecule in a hydrated DPPC/cholesterol (1:1) lipid bilayer at 303.15 K (L$_o$ state) for 500 ns of MD production phase.

To obtain insights on the DOX$^+$ location, we calculate the electron density profile for the main groups in the DOX$^+$/DPPC$^{Lo}$/cholesterol system along the z-direction to the bilayer normal over the last 100 ns of MD production phase (Figure 4).

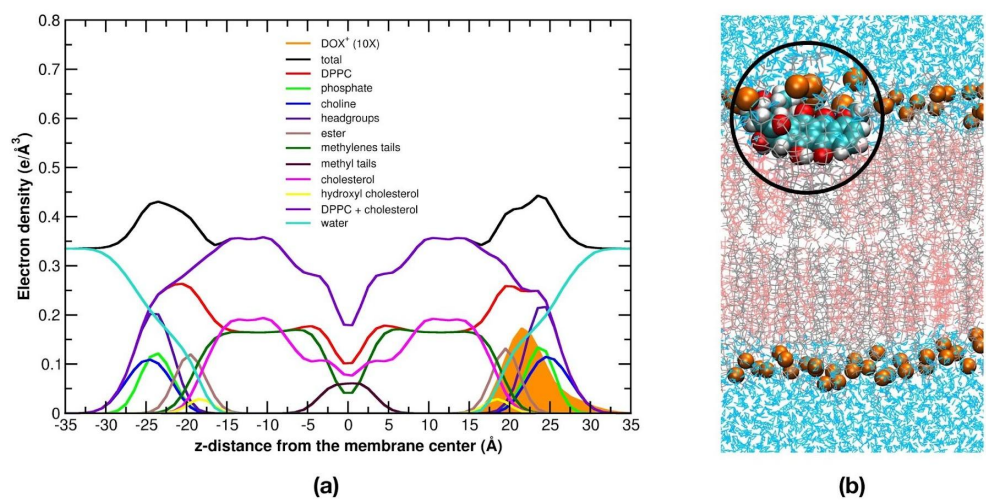

Figure 4. (a) Electron density profiles of the main lipid groups in DOX$^+$/DPPC$^{Lo}$/cholesterol system over the last 100 ns of MD production phase. (b) Final snapshot from the MD production phase. Water molecules are shown in blue, DPPC lipids in grey and DPPC phosphate groups in orange. For the sake of clarity, the intensity of the DOX$^+$ electron density profile is enhanced by 10 times.

In Figure 4, we notice the DOX$^+$ location at the interfacial region between the DPPC-PO$_4^-$ and DPPC-ROR$^|$ groups in the DPPC$^{Lo}$/cholesterol (1:1) bilayer. We identify the maximum peak of DOX$^+$ profile showing up at about 21.5 Å from the bilayer center, where it might establish favorable interactions with the polar groups in the DPPC$^{Lo}$/cholesterol bilayer interface. Hence, to better understand the molecular interactions that favor the DOX$^+$ penetration from an energetic perspective, we track the electrostatic and vdW non-bonded interactions for both DOX$^+$/DPPC and DOX$^+$/cholesterol pairs (Figure 5).

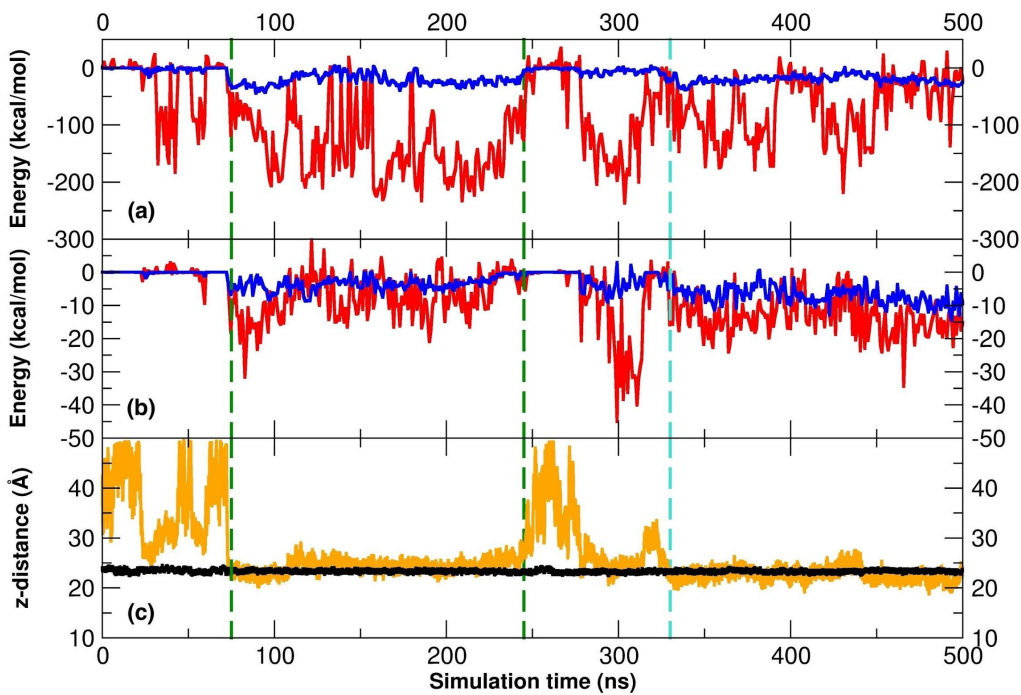

Figure 5. (a,b) Electrostatic (red) and vdW (blue) non-bonded interaction energies between (a) $DOX^+$/DPPC and (b) $DOX^+$/cholesterol. (c) z-component of C.O.M distance of $DOX^+$ (orange) and $DPPC-PO_4^-$ groups (black) from the bilayer center in $DOX^+$/$DPPC^{Lo}$/cholesterol system over 500 ns MD production phase.

Figure 5 shows that $DOX^+$ starts to penetrate the $DPPC^{Lo}$/cholesterol headgroups region within 100-200 ns of MD production phase, evidenced by lower values in the vdW and electrostatic interaction energies between $DOX^+$ and DPPC or cholesterol molecules (Figures 5a and 5b). Moreover, we notice that the C.O.M. of $DOX^+$ often alternates its position along the z-direction between the bulk-water phase and the $DPPC-PO_4^-$ headgroups in the first 350 ns of MD production phase. From this simulation point on, the z-distance plot (Figure 5c) shows that $DOX^+$ settles itself at the interface between the $DPPC-PO_4^-$ and $DPPC-ROR^|$ groups in the $DPPC^{Lo}$/cholesterol (1:1) bilayer, in line with the EDP results in Figure 4.

Although the energetic analysis above gives a clear picture of the overall interactions between drug/membrane, no information regarding the molecular nature of these interactions can be accessed. Hence, to precisely characterize the molecular vicinity around DOX$^+$ and the molecular nature of the non-bonded interactions between DOX$^+$ and lipids, we compute the RDF and the H-bond formation involving the polar groups of DOX$^+$ and the main polar moieties of DPPC and cholesterol molecules, as well as the drug/water interactions, over the last 100 ns of MD production phase (Figure 6).

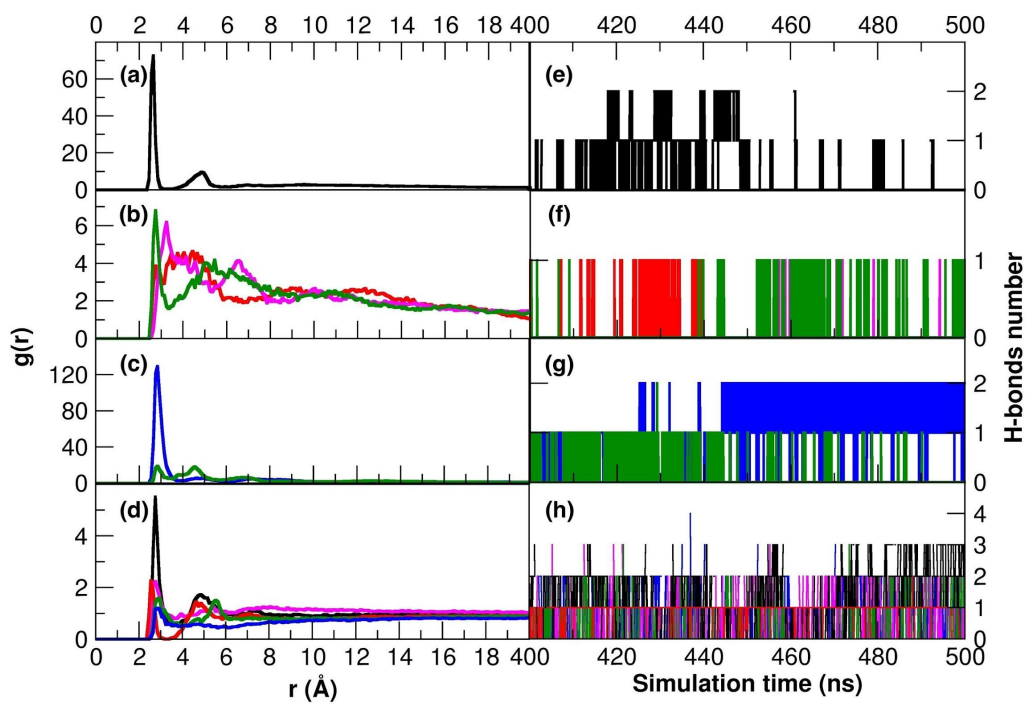

Figure 6. (a,b,c) Radial distribution function and (e,f,g) number of H-bond formation for the DOX-NH$_3^+$/DPPC-PO$_4^-$ (black), DOX-O4H/DPPC-ROR$^|$ (red), DOX-O6H/DPPC-ROR$^|$ (magenta), DOX-O8H/DPPC-ROR$^|$ (green), DOX-O3H/Chol-OH (blue) and DOX-O8H/Chol-OH (green) pairs in the DOX$^+$/DPPC/Chol system over the last 100 ns MD production phase. (d) Radial distribution function and (h) number of H-bond formation for the DOX-NH$_3^+$/water (black), DOX-O1H/water (red), DOX-O4H/water (green), DOX-O8H/water (blue) and DOX-O9H/water (magenta) pairs in the DOX$^+$/DPPC/Chol system over the last 100 ns MD production phase.

The DOX-NH$_3^+$ group preferentially interacts with the oxygen atoms of DPPC-PO$_4^-$ (Figures 6a and 6d), whereas the oxygen atoms O3, O4, O6 and O8 of DOX$^+$ (see Scheme 1 for atoms labeling) favor the H-bond formation with the DPPC-ROR$^|$ groups. However, we observe that the interaction between the oxygen atom O3 of DOX$^+$ and Chol-OH groups is the most favorable among all, therefore, showing the highest RDF peak intensity and number of H-bond formation over the 500 ns of MD production phase (Figure 6c). This latter observation is a direct consequence of DOX$^+$ placing itself at the interface between the DPPC-PO$_4^-$ and Chol-OH groups, where the drug can establish multiple H-bonds with these polar lipid groups, in line with the EDP findings in Figure 4.

Regarding DOX$^+$-water interactions, the nitrogen and the O1, O4 and O8 atoms of DOX$^+$ are once again the ones interacting the most with water, as shown in Figure 6d and Figure 6h. In this case, also the carbonyl oxygen O9 has significant interactions with the solvent.

**3.2. PSM bilayers**

*3.2.1. Doxorubicin in pure PSM bilayers*

In this section, we consider the case study of a single DOX$^+$ molecule in PSM bilayers either in a solid-ordered (PSM$^{So}$) or a liquid-disordered (PSM$^{Ld}$) lipid phase through a total 500 ns of MD simulation time. To assess the DOX$^+$ location in pure PSM$^{So}$ and PSM$^{Ld}$ bilayers, we calculate the electron density profile along the bilayer normal direction over the last 100 ns of MD production phase. Figure 7 shows the electron density profile for the DOX$^+$ molecule and the main lipid groups of PSM lipids.

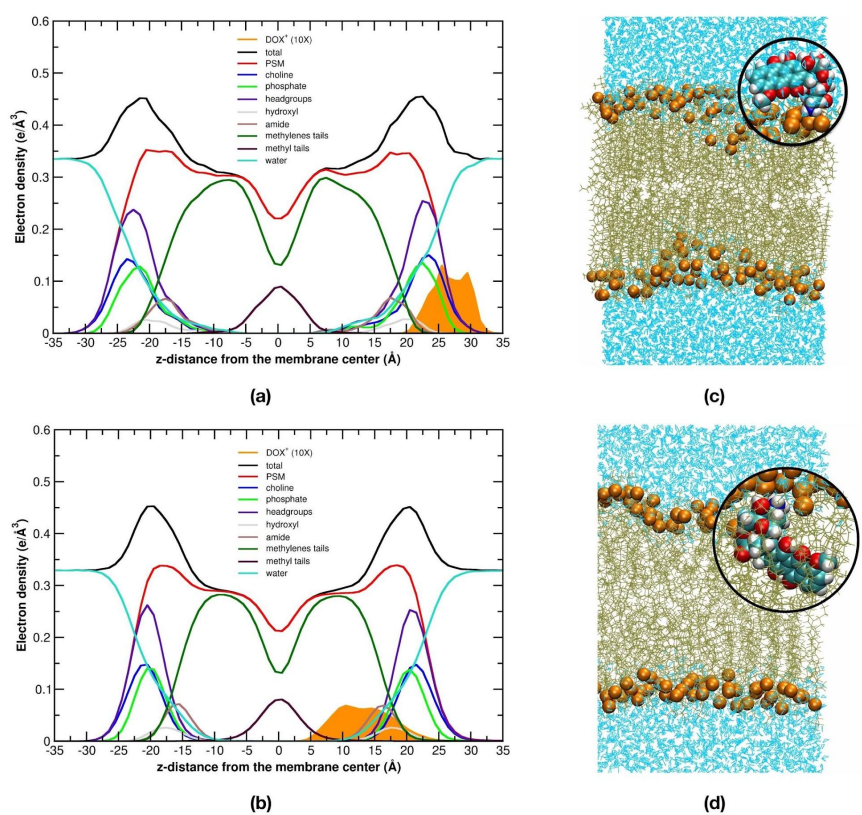

Figure 7. (a,b) Electron density profiles of the main lipid groups in (a) DOX$^+$/PSM$^{So}$ and (b) DOX$^+$/PSM$^{Ld}$ systems over the last 100 ns MD production phase. (c,d) Final snapshot from the MD production phase of (c) DOX$^+$/PSM$^{So}$ and (d) DOX$^+$/PSM$^{Ld}$ systems. Water molecules are shown in blue, PSM lipids in tan and PSM phosphate groups in orange. For the sake of clarity, the intensity of the DOX$^+$ electron density profile is enhanced by 10 times.

Analysis of Figure 7a indicates that DOX$^+$ is settled at the membrane/water interface in the DOX$^+$/PSM$^{So}$ system, with its rigid backbone lying in the water phase, whereas in the DOX$^+$/PSM$^{Ld}$ system (Figure 7b) the drug is completely immersed in the lipid phase, mainly driven by increased hydrophobic interactions between DOX$^+$ and the PSM$^{Ld}$ lipids. To investigate in depth this last assumption and shed light on the DOX$^+$/PSM interactions from an energetic perspective, we calculate the electrostatic and vdW terms between DOX$^+$ and the PSM$^{So}$ and the PSM$^{Ld}$ lipids, as shown in Figure 8, which also includes the temporal evolution of the z-component of C.O.M. distance of $DOX^+$ and $PSM-PO_4^-$ headgroups from the bilayer center.

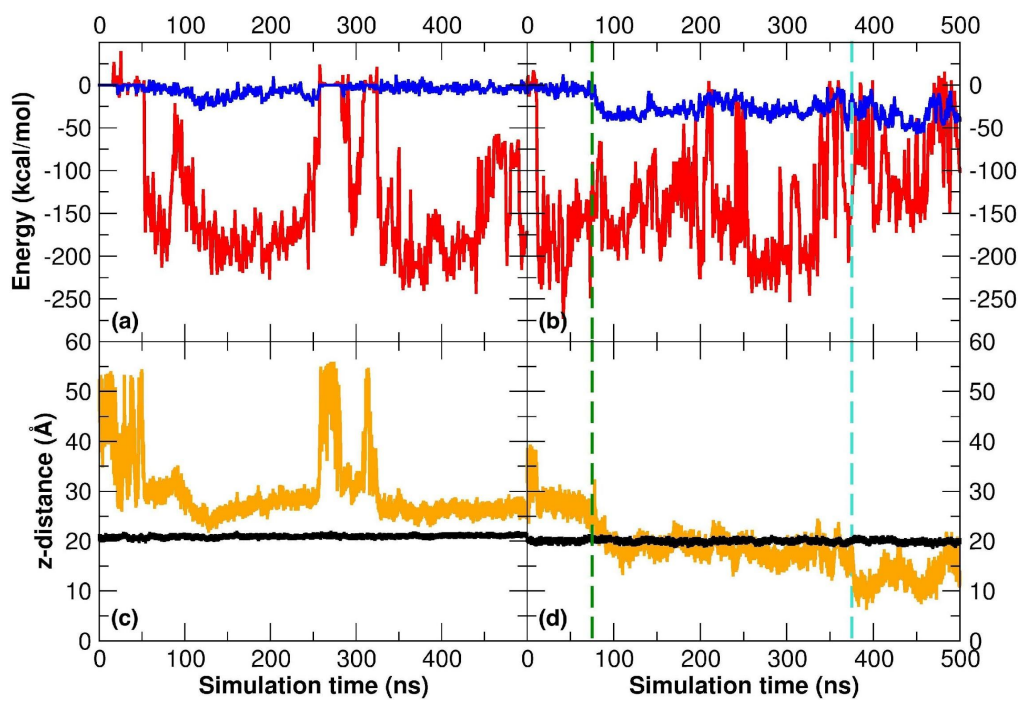

Figure 8. (a,b) Electrostatic (red) and vdW (blue) non-bonded interactions for the (a) $DOX^+/PSM^{So}$ and (b) $DOX^+/PSM^{Ld}$ systems. (c,d) z-component of C.O.M. distance of $DOX^+$ (orange) and $PSM-PO_4^-$ (black) groups from the bilayer center for the (c) $DOX^+/PSM^{So}$ and (d) $DOX^+/PSM^{Ld}$ systems over 500 ns of MD production simulation.

Based on the analysis of the results in Figure 8, we conclude that $DOX^+$ is particularly stabilized within the PSM bilayer in the liquid-disorder lipid phase. Indeed, the $DOX^+$ penetration into the $PSM^{Ld}$ bilayer is aided by the higher lipid disorder in this system, leading to an enhanced non-bonded vdW and electrostatic stabilization. The electrostatic term has an average value at about 10-fold lower than the vdW term, therefore, mainly contributing to the $DOX^+$ stabilization in the pure $PSM^{Ld}$ and $PSM^{So}$ bilayers. Furthermore, Figures 8c and 8d display that $DOX^+$ penetrates deeper the PSM$^{Ld}$ than the PSM$^{So}$ bilayer, confirming the EDP findings in Figure 7.

Undoubtedly, the preferred location of DOX$^+$ depends on its molecular vicinity and the nature of the interactions established there. To further investigate the molecular vicinity around DOX$^+$ and characterize the most relevant H-bond formations that contribute to the enhanced stabilization of DOX$^+$ in PSM bilayers, we analyze these parameters over the last 100 ns MD production phase in Figure 9. The drug/water interactions are studied, as well.

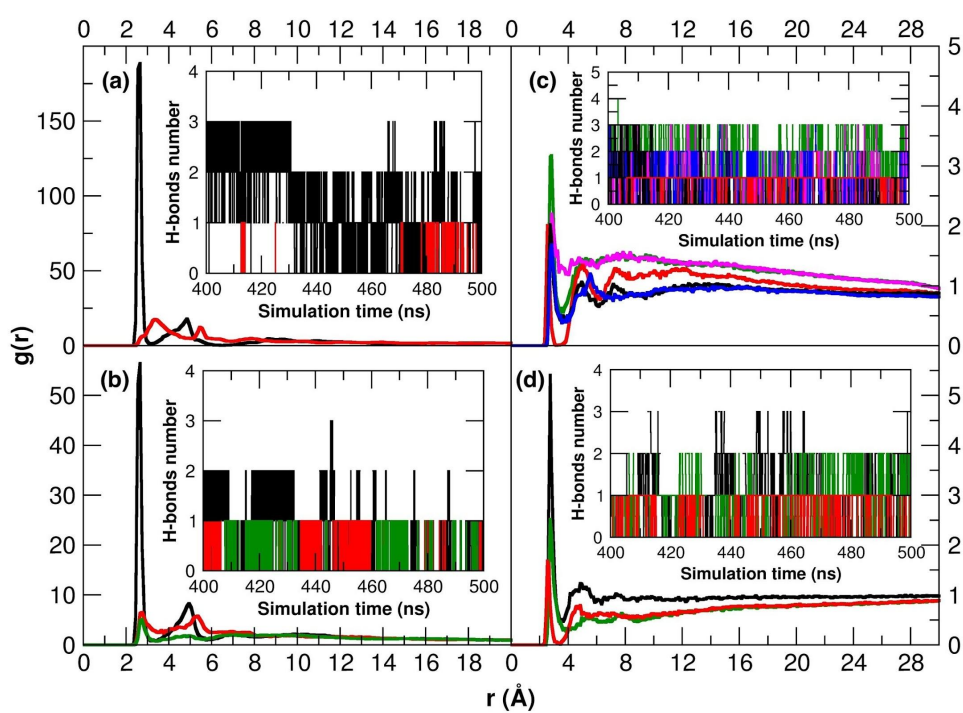

Figure 9. (a, b) Radial distribution function and number of H-bond formation (insets) for the DOX-NH$_3^+$/PSM-PO$_4^-$ (black), DOX-O4H/PSM-PO$_4^-$ (red) and DOX-O8H/PSM-PO$_4^-$ (green) pairs in (a) pure PSM$^{So}$ and (b) PSM$^{Ld}$ systems over the last 100 ns MD production phase. (c, d) Radial distribution function and number of H-bond formation (insets) for the DOX-NH$_3^+$/water (black), DOX-O1H/water (red), DOX-O3H/water (red), DOX-O4H/water (blue) and DOX-O8H/water (red) pairs in (c) pure PSM$^{So}$ and (d) PSM$^{Ld}$ systems over the last 100 ns MD production phase.

Figure 9 shows that the DOX-NH$_3^+$ group and, to a lesser extent, the DOX-O4H group (please see Scheme 1a for atoms labeling), establish H-bonds with the PSM-PO$_4^-$ group independently on the PSM lipid phase. We also find a similar trend in the RDF distributions calculated between the main polar groups of DOX$^+$ and the PSM-PO$_4^-$ groups in the PSM$^{Ld}$ and PSM$^{So}$ systems. However, there is a decrease in the former system (Figure 9b), compared with the latter one (Figure 9a), in the number of PSM-PO$_4^-$ groups surrounding the DOX-NH$_3^+$ group. We observe that the So lipid phase favors the H-bond formation between the DOX-NH$_3^+$ group and the PSM-PO$_4^-$ groups (Figure 9a, inset), being this number considerably lower in the DOX$^+$/PSM$^{Ld}$ system (Figure 9b, inset).

The DOX$^+$-water interaction patterns of Figure 9c and Figure 9d resemble the ones for the DPPC$^{So}$ and DPPC$^{Ld}$ cases. Indeed, the extent of hydrogen bonds is a way higher and heterogeneous in the PSM$^{So}$ case, where DOX$^+$ nitrogen, O1, O3, O4 and O8 atoms (please see Scheme 1a for atoms labeling) establish H-bonds with the solvent, rather than in the PSM$^{Ld}$ case, where only N, O1 and O3 do.

Based on the above, the interplay between molecular interactions and dynamics of DOX$^+$ in the PSM$^{So}$ bilayer can be summarized as follows: within the first 100 ns of MD production phase, DOX$^+$ is hauled from the bulk water towards the PSM$^{So}$ headgroups region and, then, it starts to alternate its interaction with the PSM-PO$_4^-$ groups and the bulk water in the subsequent 300 ns. Once DOX$^+$ reaches the PSM headgroups region, we observe that DOX$^+$ C.O.M. remains stable at a z-distance of about 25.5 Å from the bilayer center. We notice that DOX$^+$ mainly establishes its interactions with the PSM$^{So}$ bilayer through H-bonds between the DOX-NH$_3^+$ group and the PSM-PO$_4^-$ headgroups.

### 3.2.2. Doxorubicin in cholesterol-enriched PSM bilayer

In this section, we investigate the influence of cholesterol on the DOX$^+$ interactions with a binary mixture of PSM lipids enriched with 50% cholesterol. To this end, we analyze a single DOX$^+$ molecule in the presence of a hydrated PSM/cholesterol (1:1) lipid bilayer at 303.15 K (L$_o$ state) for 500 ns of MD production phase.

To estimate the DOX$^+$ location in the PSM$^{Lo}$/cholesterol (1:1) bilayer, we calculate the EDP along the bilayer normal direction for DOX$^+$ and the main lipid groups over the last 100 ns of MD production phase (Figure 10).

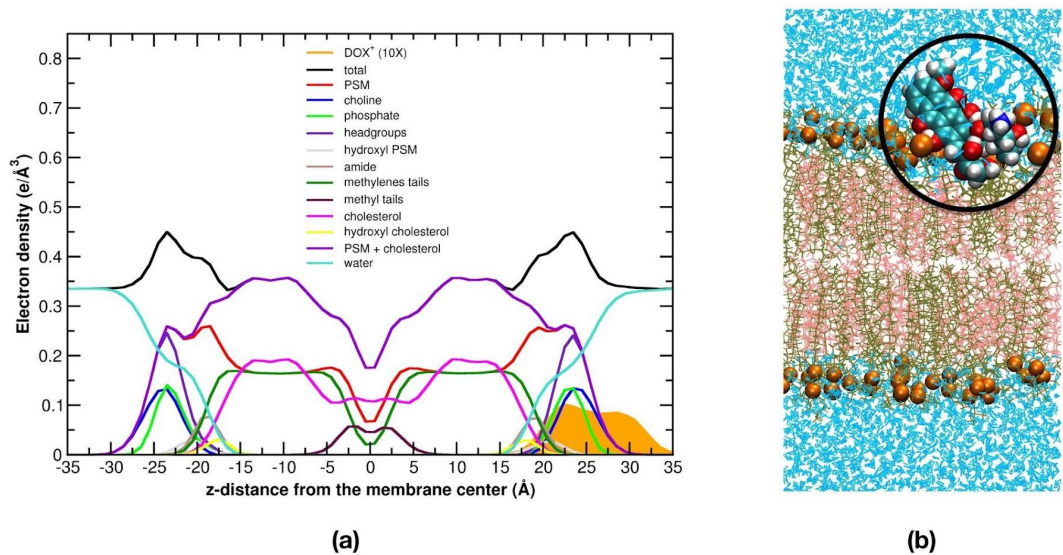

Figure 10. (a) Electron density profiles of the main lipid groups in the DOX$^+$/PSM$^{Lo}$/cholesterol system over the last 100 ns of MD production simulation. (b) Final snapshot from the MD production phase. Water molecules are shown in blue, PSM lipids in tan and PSM phosphate groups in orange. For the sake of clarity, the intensity of the DOX$^+$ electron density profile is enhanced by 10 times.

Figure 10 shows that the maximum EDP peak of DOX$^+$ is settled at a z-distance of about 22 Å from the bilayer center, within the bilayer region that comprehends the PSM-PO$_4^-$ and PSM-OH groups. Furthermore, we also identify a smaller shoulder at about 28 Å from the bilayer center, located at the lipid/water interface. These results indicate that DOX$^+$ has at least two metastable locations along the bilayer normal. One is deeper in the bilayer, located within the headgroups region, while the second one takes place outside the bilayer, at the membrane-water interface. These two metastable locations of DOX$^+$ are intimately correlated to the H-bond interactions established and the molecular vicinity around the drug. Hence, to further investigate the interplay of DOX$^+$ with the main polar groups in the PSM/cholesterol (1:1) bilayer, we evaluate the RDF and H-bond formations between DOX$^+$ and the polar moieties of PSM lipids and the water molecules. Figure 11 shows the RDF profiles and H-bond formation (insets) between the DOX-$NH_3^+$ and the oxygen atoms of PSM-$PO_4^-$ and carbonyl groups of PSM (from now on PSM-NC=O), together with the drug/water interactions.

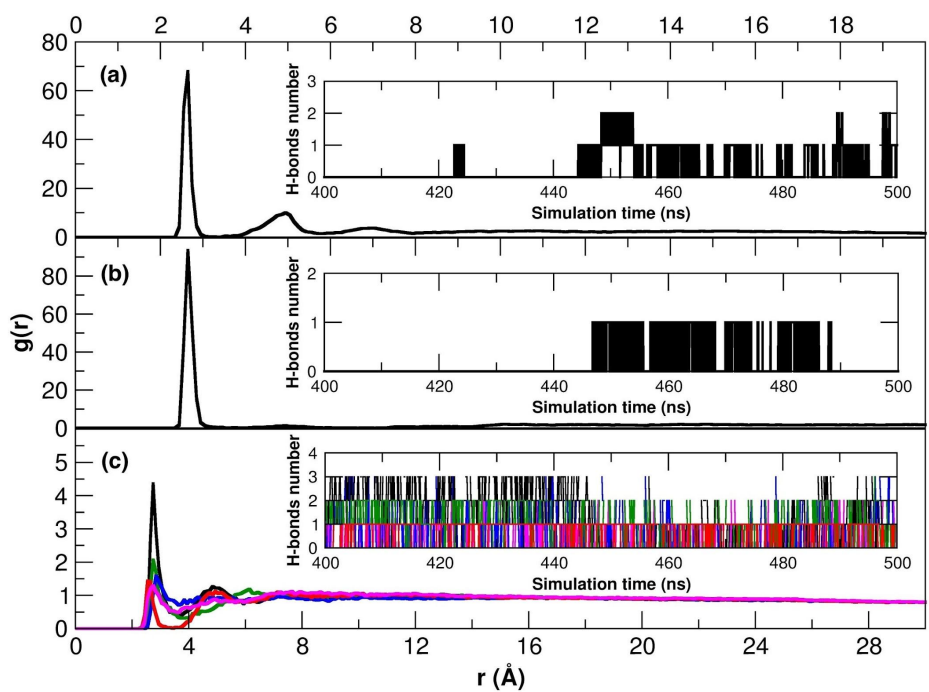

Figure 11. (a, b) Radial distribution function and number of H-bond formation (insets) for the (a) DOX-$NH_3^+$/PSM-$PO_4^-$ and (b) DOX-$NH_3^+$/PSM-NC=O pairs in $PSM^{Lo}$/cholesterol system over the last 100 ns of MD production simulation. (c) Radial distribution function and number of H-bond formation (inset) for DOX-$NH_3^+$/water (black), DOX-O1H/water (black), DOX-O3H/water (red), DOX-O8H/water (green) and DOX-O9H/water (magenta) pairs in $PSM^{Lo}$/cholesterol system over the last 100 ns of MD production phase.

The examination of Figure 11 shows that the DOX-$NH_3^+$ group is mainly surrounded by the polar groups PSM-NC=O and PSM-$PO_4^-$, with the former pair (Figure 11b) having a slightly higher RDF intensity compared to the latter one (Figure 11a). In the last 50 ns of MD production phase, we observe the absence of H-bonds between the DOX-$NH_3^+$ group and the PSM-NC=O groups, while the H-bond formation remains practically unchanged between the DOX-$NH_3^+$/PSM-$PO_4^-$ pair. As for the DOX$^+$-water interactions, the drug establishes hydrogen bonds with the solvent molecules through its N, O1, O3, O8 and O9 atoms (Figure 11c), a situation that is very close to the DPPC/Chol$^{Lo}$ case.

To better understand these interactions from an energetic point of view, we examine the non-bonded interactions and the z-component of the C.O.M. distance of DOX$^+$ and the PSM-PO$_4^-$ in the PSM/cholesterol system. Figure 12 displays the vdW and electrostatic contributions to the total non-bonded energy for the DOX$^+$/PSM and DOX$^+$/cholesterol pairs computed over the 500 ns of MD production phase.

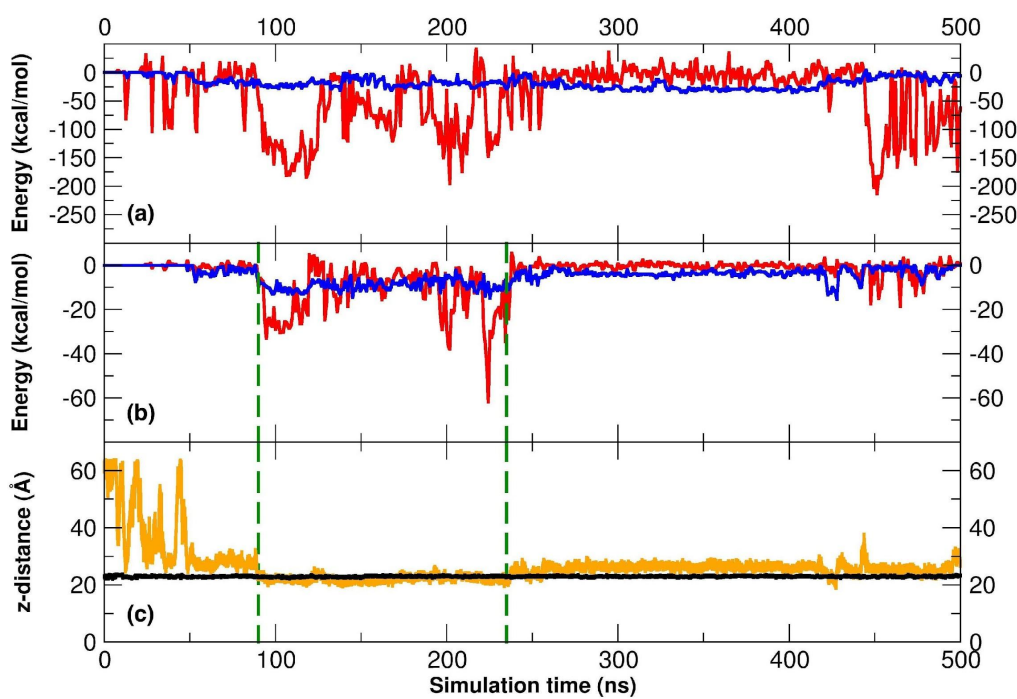

Figure 12. (a,b) Electrostatic (red) and vdW (blue) non-bonded interaction energy for (a) DOX$^+$/PSM$^{Lo}$ and (b) DOX$^+$/cholesterol pairs. (c) z-component of C.O.M distance of DOX$^+$ (orange) and PSM-PO$_4^-$ groups (black) from the bilayer center over 500 ns of MD production phase.

We see that the electrostatic interactions between DOX$^+$ and the PSM and cholesterol lipids considerably lowers its values when the drug reaches the membrane-water interface at about 100 ns of MD production phase (Figures 12a and 12b). This energetic stabilization, which comes from membrane/drug interactions, persists and becomes even more intense up to 250 ns MD production phase. From this point on, we identify that interactions in the $DOX^+$/PSM and $DOX^+$/Chol pairs become weaker, evidenced by a substantial increase in the average value of non-bonded energies within the time range of 300-450 ns of MD production phase. In the last 50 ns of MD production, $DOX^+$ establishes interactions with the PSM bilayer once again, specifically between $DOX^+$ and the PSM lipids, leading to a further energetic drop in the electrostatic interactions.

Based on the results above, the $DOX^+$ penetration process in the $PSM^{Lo}$/cholesterol can be understood as follows: at about 100 ns MD production simulation, $DOX^+$ approaches the water/lipid interface, dragged from the bulk water, where it establishes preferential interactions with $PSM-PO_4^-$ groups. Up to 250 ns of MD production, $DOX^+$ penetrates the headgroups region settling itself at a z-distance of 20 Å from the bilayer center, keeping its rigid backbone positioned between the hydrocarbon chains and the lipid headgroups. From this simulation point on, $DOX^+$ starts to lose the interactions between its polar groups belonging to the rigid backbone and the hydroxyl groups of both cholesterol and PSM lipids, resulting in a flip of the rigid backbone along the z-direction towards the bulk water, where it holds stable up to 500 ns of MD production.

### *3.2.3. Doxorubicin translocation in pure and cholesterol-enriched PSM bilayers*

To investigate the thermodynamic aspects of drug translocation in different PSM lipid phases, namely pure $PSM^{So}$ and $PSM^{Lo}$/cholesterol lipid bilayers, we calculate the total free energy of $DOX^+$ partitioning from the bulk water to relevant bilayer regions. To this end, we adopt the thermodynamic cycle shown in Scheme 2.

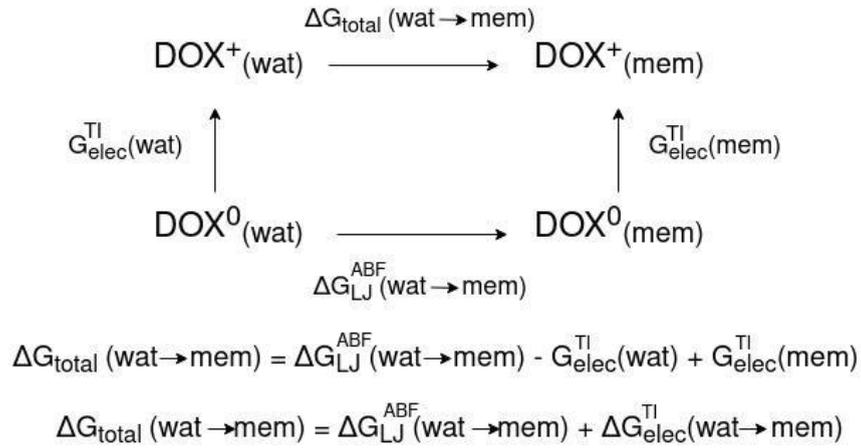

$$\Delta G_{total}(\text{wat} \rightarrow \text{mem}) = \Delta G_{LJ}^{ABF}(\text{wat} \rightarrow \text{mem}) - G_{elec}^{TI}(\text{wat}) + G_{elec}^{TI}(\text{mem})$$

$$\Delta G_{total}(\text{wat} \rightarrow \text{mem}) = \Delta G_{LJ}^{ABF}(\text{wat} \rightarrow \text{mem}) + \Delta G_{elec}^{TI}(\text{wat} \rightarrow \text{mem})$$

Scheme 2. Thermodynamic cycle adopted in this work to estimate the total free energy of DOX$^+$ partitioning in PSM bilayers. DOX$^0$ stands for the "uncharged" DOX molecule whereas DOX$^+$ represents its "fully-charged" state.

According to the scheme, we decompose the total free energy of DOX$^+$ into its polar and nonpolar components, allowing a more detailed analysis of how each of these terms contributes to DOX$^+$ partitioning in the main lipid regions of pure and cholesterol-rich PSM bilayers. Scheme 3 shows the membrane regions and their main lipid moieties, for which the free energy differences of DOX$^+$ partitioning were estimated.

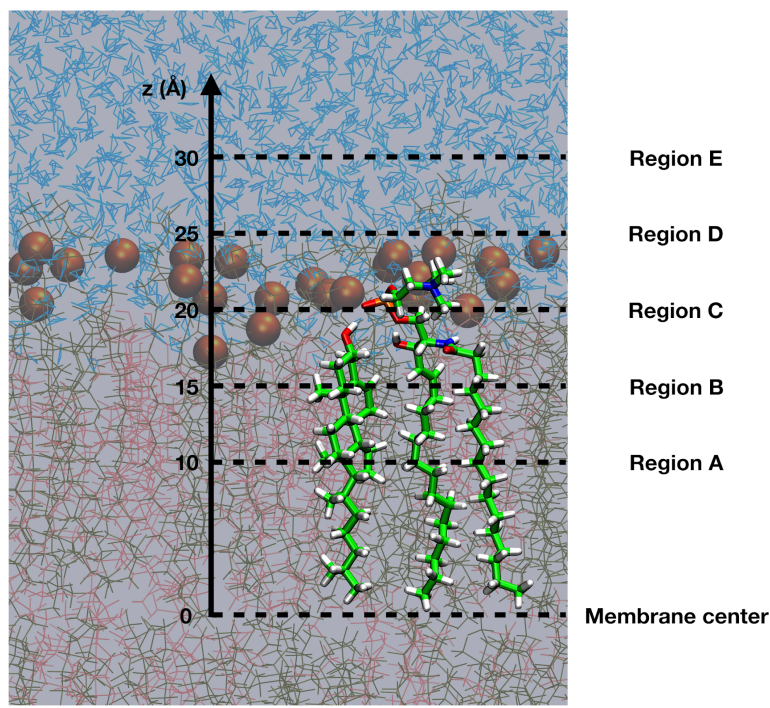

Scheme 3. Graphical representation of the membrane regions (A-E) adopted to evaluate the free energies of DOX$^+$ partitioning in PSM$^{So}$ and PSM$^{Lo}$/cholesterol (1:1) bilayers. In the foreground, a cholesterol (left) and a PSM (right) molecule are represented: carbon is shown in green, hydrogen in white, oxygen in red, nitrogen in blue and phosphorus in orange. In the background, water molecules are shown in blue, PSM lipids in tan, cholesterol in pink and PSM phosphate groups in orange.

To this end, we first translocate an "uncharged" DOX molecule (from now on DOX$^0$), where all its partial atomic charges are set to zero, to obtain the LJ free energy profiles (Figure S1), and, then, the LJ free energy of DOX$^0$ partitioning from the bulk water towards the lipid bilayer core ($\Delta G_{LJ}^{wat \to mem}$) via ABF calculations.[56-59] We estimate the electrostatic work to charge a DOX$^+$ molecule harmonically constrained in relevant bilayer regions via TI calculations[60-63] to assess the electrostatic free energy of DOX$^+$ partitioning ($\Delta G_{elec}^{wat \to mem}$). The protocol details of the free energy calculations and a detailed discussion on the limitations of these methods can be found in the Supplementary Material. Tables 1a and 1b show the total free energy of DOX$^+$ partitioning for relevant regions in the pure PSM$^{So}$ and PSM$^{Lo}$/cholesterol bilayers, decomposed in their LJ and electrostatic free energy contributions.

Table 1a. Free energies of DOX$^+$ partitioning in pure PSM$^{So}$ bilayer

| (kcal/mol) | Region A | Region B | Region C | Region D | Region E |
|---|---|---|---|---|---|
| $\Delta G^{LJ}_{wat \to mem}$ | -12.2 (± 2.3) | -4.7 (± 2.4) | 6.0 (± 2.4) | 5.7 (± 2.6) | 0.5 (± 2.6) |
| $\Delta G^{elec}_{wat \to mem}$ | 5.7 (± 2.5) | 5.1 (± 1.5) | 1.6 (± 1.8) | -4.1 (± 3.0) | -8.2 (± 3.4) |
| $\Delta G^{total}_{wat \to mem}$ | -6.5 (± 3.4) | 0.4 (± 2.8) | 7.6 (± 3.0) | 1.6 (± 4.0) | -7.7 (± 4.3) |

Table 1b. Free energies of DOX$^+$ partitioning in PSM$^{Lo}$/cholesterol (1:1) bilayer

| (kcal/mol) | Region A | Region B | Region C | Region D | Region E |
|---|---|---|---|---|---|
| $\Delta G^{LJ}_{wat \to mem}$ | -6.8 (± 2.8) | -4.5 (± 2.8) | 0.8 (± 3.0) | 0.3 (± 3.1) | 0.4 (± 3.2) |
| $\Delta G^{elec}_{wat \to mem}$ | 26.9 (± 3.6) | 7.2 (± 1.2) | -7.7 (± 1.7) | -8.8 (± 2.8) | -8.0 (± 1.1) |
| $\Delta G^{total}_{wat \to mem}$ | 20.1 (± 4.6) | 2.7 (± 3.0) | -6.9 (± 3.5) | -8.5 (± 4.2) | -7.6 (± 3.4) |

The comparison of the total free energy difference of DOX$^+$ partitioning in Tables 1a and 1b reveals that cholesterol favors the drug partitioning into a broader hydrophilic region of the PSM$^{Lo}$/cholesterol bilayer (Regions C, D and E) compared to the pure PSM$^{So}$ bilayer. In this latter system, we observe that DOX$^+$ partitioning is spontaneous at the PSM$^{So}$/bulk-water interface (Region E) and has unfavorable partitioning at the hydrophilic/hydrophobic transition regions (Regions C and D), showing an opposite trend to what we observe in the presence of cholesterol.

Regarding the LJ free energy difference to translocate DOX$^0$ across the lipid bilayer, namely LJ free energy of DOX$^0$ partitioning, we register a similar trend in both pure PSM$^{So}$ and PSM$^{Lo}$/cholesterol systems, where the thermodynamic process of DOX$^0$ transfer from bulk water into hydrophobic regions of the membrane (Regions A and B) is spontaneous. In contrast, the LJ free energy of DOX$^0$ partitioning is thermodynamically unfavorable into hydrophilic regions of these lipid bilayers (Regions C, D and E), as one would expect. We observe that the LJ free energy cost to displace $DOX^0$ from the bulk water across the lipid headgroups (Figure S1) is higher in the $PSM^{So}$ bilayer (~ 8 kcal/mol) than that in the presence of cholesterol (~ 2 kcal/mol). Once $DOX^0$ overcomes this hydrophobic barrier, the LJ free energy of $DOX^0$ partitioning into the apolar environment of the $PSM^{So}$ bilayer (Region A) is a more favorable thermodynamic process (~ 5 kcal/mol), if compared to the same region of the $PSM^{Lo}$/cholesterol bilayer. In contrast, as one would expect, we find unfavorable electrostatic free energy partitioning of $DOX^+$ into the bilayers' hydrophobic regions. We identify that the electrostatic free energy of $DOX^+$ partitioning is thermodynamically favorable into hydrophilic regions of the $PSM^{So}$ (Regions D and E) and $PSM^{Lo}$/cholesterol (Regions C, D and E) bilayers, whereas it is unfavorable into hydrophobic environments (Regions A and B). Interestingly, we notice that Region C does not follow the same trend in the electrostatic free energy evaluation: the electrostatic free energy of $DOX^+$ partitioning is thermodynamically favorable in the presence of cholesterol while it is unfavorable in its absence.

## 4. Discussion

### *4.1. Liquid-disordered lipid phase favors $DOX^+$ penetration in pure DPPC and PSM bilayers.*

Experimental findings have shown that DOX insertion is facilitated in less condensed PC monolayers[64], and its localization within membranes mainly depends on the molecular packing of lipids[65]. For instance, Tritton et al.[66] have found enhanced penetration of DOX into PC liposome membranes in the liquid-crystalline phase. In contrast, they noticed that a deeper membrane penetration of DOX is impaired in the gel phase. In line with these experimental findings, we also observe an enhanced penetration of $DOX^+$ when the pure DPPC and PSM lipid bilayers shift from the solid-ordered to the liquid-disordered phase (Figures 1 and 7). This result is likely a consequence of the looser packing of lipids in the liquid-disordered phase compared to the solid-ordered lipid phase. Hence, we can infer that the lipid phase regulates the penetration of lipid membranes by amphiphilic molecules such as $DOX^+$.

### *4.2. DOX$^+$ rigid backbone location is modulated by the presence of cholesterol.*

Previous experimental data [65] support the existence of two binding sites for DOX in fluid PC membranes: one is embedded in the membrane, while the other is located at the membrane-water interface. These authors argue that the former site favors hydrophobic interactions with the anthracycline moiety of DOX while the latter promotes the polar interactions between the sugar group of DOX and the lipid headgroups. Moreover, Gaber et al. [67] have observed that the addition of cholesterol to fluid PC membranes, increasing their lipid packing, decreases the overall interaction with DOX molecules. Furthermore, in a recent experimental and computational investigation, Reis et al. [32] have found a similar trend where DOX$^+$ establishes favorable interactions with the lipid headgroups region of binary PC/SM and ternary PC/SM/cholesterol lipid bilayers. They suggested that the drug location at the membrane-water interface is modulated by both electrostatic interactions between its DOX-NH$_3^+$ group and the negatively charged phosphate groups, and hydrophobic interactions between the DOX$^+$ rigid backbone and the nonpolar bilayer regions. Our results are in fair agreement with the results cited above. We noticed that DOX$^+$ has at least two stable locations in the cholesterol-rich PSM lipid bilayer (Figures 7 and 10), while the tightly packing of the pure PSM$^{So}$ membrane prevents its complete penetration. We have found that the DOX$^+$ rigid backbone dynamically alternates its location by flipping between an outer region at the PSM-water interface and an inner membrane region between the hydroxyl groups of cholesterol and the headgroups of the cholesterol-rich PSM$^{Lo}$ bilayer. The latter DOX$^+$ location is likely the most stable one in the presence of cholesterol, corroborated by the most prominent peak of DOX$^+$ at the headgroups region in the EDP profiles estimated from three independent MD simulations (Figure S13). It is noteworthy to mention that the cholesterol molecules prevent further insertion of the DOX rigid backbone into the hydrocarbon chains, as observed in the more fluid DPPC$^{Ld}$ and PSM$^{Ld}$ lipid bilayers. These observations above can be rationalized as a biophysical modulation through enhanced hydration at the headgroups region (Figure S14) and lowering of the hydrophobic free energy barrier (Figure S1) for DOX translocation across this bilayer region, caused by the replacement of PSM lipids in the solid-ordered phase by cholesterol molecules. This concerted effect that, on the one hand, favors the penetration of water molecules beyond the headgroups towards the membrane center, and, on the other hand, decreases the hydrophobic free energy barrier to translocate DOX across the headgroups region, facilitates the DOX$^+$ rigid backbone translocation through the PSM/cholesterol headgroups region.

*4.3. Cholesterol decreases the overall interactions between DOX$^+$ and lipids.*

In cholesterol-enriched DPPC$^{Lo}$ and PSM$^{Lo}$ bilayers, vdW interactions between DOX$^+$ and lipids are more stable than those in the solid-ordered lipid state (Table S2). This apolar stabilization is counterbalanced by less favorable electrostatic interactions between DOX$^+$ and lipids. Furthermore, we observe an overall loss of non-bonded interactions between DOX$^+$ and lipids due to the presence of cholesterol in binary DPPC$^{Lo}$ and PSM$^{Lo}$ bilayers. The results above are in line with those of Reis et al. [32], who reported that the presence of cholesterol decreased the extent of non-bonded interactions between DOX and PC lipids. We rationalize this loss of non-bonded interactions between DOX$^+$ and lipids due to increased accessibility of water molecules into the headgroups region in the presence of cholesterol (Figure S14), therefore, competing with lipids to interact with the drug.

*4.4. DOX$^+$ partitioning is spontaneous into the headgroups region of PSM/cholesterol bilayer.*

DOX$^+$ partitioning is thermodynamically favorable into the hydrophilic regions of both pure PSM$^{So}$ and cholesterol-rich PSM$^{Lo}$ bilayer systems (Tables 1a and 1b). Previous study has addressed similar free energy analyses for the translocation of DOX across pure and cholesterol-rich DPPC bilayers[31], where DOX partitioning is spontaneous into the headgroups region of these membranes. These authors also observed an overall decrease of DOX partitioning into the membrane core upon addition of cholesterol. Furthermore, Shinoda et al. [68] have found a higher free energy barrier to translocate a water molecule into the hydrophobic core of PSM bilayers due to the presence of cholesterol. These findings above align with our predictions in Tables 1a and 1b, where we also found a higher energetic barrier to translocate DOX$^+$ into the cholesterol-rich PSM bilayer core. Furthermore, Reis et.al.[32] concluded, through a combined experimental and MD simulation study, that DOX$^+$ has a similar partitioning into PC/SM liposomes in the presence or absence of cholesterol. Despite the different composition and lipid phase compared to our PSM bilayer models, further analysis of Tables 1a and 1b shows that, indeed, there is a common region (Region E) at the lipid-water interface in which DOX$^+$ has a favorable partitioning in the pure PSM$^{So}$ and the cholesterol-rich PSM$^{Lo}$ bilayers. However, we observe a broader hydrophilic region in the latter system (Regions C, D and E) in which DOX$^+$ has a favorable thermodynamic partitioning, whereas it is limited to the lipid-water interface in the former one (Region E).

## 5. Conclusions

Many research efforts are devoted to improving the efficiency of chemotherapy. One successful application is the transportation of drugs by embedding the therapeutics into lipid carriers. However, it is also essential to consider how drug entrapment in these carrier membranes can influence their pharmacological activity. In this context, one of the most significant models to investigate this issue is portrayed by DOX and liposome membranes.

Our present study clarifies how different membrane compositions and lipid phases affect the DOX$^+$ location and its interaction with the polar lipid region of these membranes. Based on a systematic comparative analysis of DOX$^+$ interactions with two major classes of lipid membranes, composed of DPPC and PSM in the presence and absence of cholesterol, our results demonstrate that DOX$^+$ penetration is favored in less ordered lipid phases. By changing the lipid phase and modifying the molecular nature in the polar region of PSM lipid bilayers, cholesterol can also modulate the DOX$^+$ penetration and partitioning in these membranes. Our free energy results support that DOX$^+$ partitioning is favored into a broader hydrophilic region of the cholesterol-rich PSM$^{Lo}$ membrane, whereas it is more localized at the lipid-water interface in the pure PSM$^{So}$ membrane.

These findings provide an atomistic insight into how lipid composition and membrane fluidity influence the liposome's membrane ability to load DOX. This valuable information could be translated into an enhanced therapeutic efficacy of liposomal anticancer drug products by optimizing their formulations. In addition, we envision that the computational framework adopted here, further combined with constant-pH MD methods, paves the way to future investigations of drug release mechanisms, such as those involved in pH-responsive membranes and pH-sensitive controlled drug release.

**Acknowledgments**

The project has received funding from the European Research Council (ERC) under the European Union's HORIZON2020 research and innovation programme (ERC Grant Agreement No [647020]). This work has been partially realized in the context of the project NEVERMIND "Nuove frontiere nello sviluppo di nanofarmaci per il miglioramento dell'efficacia e della sicurezza terapeutica nelle patologie neurologiche", CP2_16/2018 – Collaborative project II edition FRRB (Fondazione Regionale per la ricerca biomedica).

# Supplementary Material: "Molecular Dynamics simulations of Doxorubicin in sphingomyelin-based lipid membranes"

Paulo Siani,[1] Edoardo Donadoni,[1] Lorenzo Ferraro,[1] Francesca Re,[2,3] Cristiana Di Valentin[1,3]

[1] Department of Materials Science, University of Milano-Bicocca
via R. Cozzi 55, 20125 Milano Italy

[2] School of Medicine and Surgery, University of Milano-Bicocca,
via Raoul Follereau 3, Vedano al Lambro, MB 20854, Italy

[3] BioNanoMedicine Center NANOMIB, University of Milano-Bicocca

# 1. Computational details

## 1.1. Free energy calculations

To improve the free energy convergence, and therefore, the accuracy of these calculations, we adopted a combination of free energy methods to, separately, calculate the energetic components that constitute the total free energy of $DOX^+$ partitioning into PSM membranes. An important outcome of this approach is to provide a better portrait of how electrostatic and non-electrostatic components contribute to the total free energy differences of drug partitioning into multiple regions in PSM membranes.

Within this framework, we use the Thermodynamic Integration and the Adaptive Biasing Force techniques to separately determine the electrostatic and the non-electrostatic contributions to translocate $DOX^+$ from the bulk-water phase to relevant bilayer depths (please see Scheme 3, main text). To obtain the total free energy of $DOX^+$ partitioning into different membrane depths, we adopted the thermodynamic cycle shown in Scheme 2 (main text). On the one hand, we aimed to mitigate the deleterious effects of singularities associated with the Lennard-Jones terms' perturbation at the end-point states, and on the other hand, provide a computationally cheap and accurate protocol to estimate the electrostatic free energy of $DOX^+$ partitioning into membranes.

## 1.2. Alchemical free energy: Thermodynamic integration (TI)

We calculated the electrostatic work of "charging" a protonated DOX molecule with the aid of the TI method implemented in the NAMD2.13 package.[1] Within the TI formalism, the electrostatic free energy difference between the thermodynamic states $\lambda = 0$ and $\lambda = 1$ can be written as

$$\Delta G^{elec}(0 \to 1) = \int_0^1 \langle \frac{\partial U^{elec}(\lambda)}{\partial \lambda} \rangle_\lambda d\lambda \quad (1)$$

where $\Delta G^{elec}$ is the electrostatic free energy difference between state $\lambda = 0$ and state $\lambda = 1$, $U^{elec}$ is the solute-solvent electrostatic potential energy as a function of the coupling parameter function $\lambda$ with integration limits from 0 to 1.

Numerical integration schemes are widely used to evaluate the above integral and have the general form

$$\Delta G^{elec}(0 \to 1) \approx \sum_{i=1}^{N} w_i \langle \frac{\partial U^{elec}(\lambda)}{\partial \lambda} \rangle_{\lambda_i} (\lambda_{i+1} - \lambda_i) \quad (2)$$

where $w_i$ are the weights and $\lambda_i$ are the thermodynamic states where the integrand is evaluated from MD simulations carried out independently. The trapezoidal rule technique has been routinely applied to approximate the definite integral (Eq. 1) in free energy calculations. However, several authors have pointed out Simpson's rule as a more accurate alternative to estimate electrostatic free energy differences when higher free-energy derivatives are included.[2,3] Indeed, further precision in the electrostatic free energy evaluation can be achieved by including intermediate states between the end-point states (e.g. $\lambda = 1/2$). This approximation has been proved accurate and a cost-effective way to estimate free energy differences of thermodynamic processes mainly driven by electrostatic interactions.[3] For the particular case where an intermediate state halfway the end-point states is considered, Simpson's rule can be expressed as follows

$$\Delta G^{elec}(0 \to 1) \approx \frac{1}{6} \left( \langle \frac{\partial U^{elec}(\lambda)}{\partial \lambda} \rangle_{\lambda=0} + \langle \frac{\partial U^{elec}(\lambda)}{\partial \lambda} \rangle_{\lambda=1} \right) + \frac{2}{3} \langle \frac{\partial U^{elec}(\lambda)}{\partial \lambda} \rangle_{\lambda=1/2} \quad (3)$$

To estimate the statistical uncertainty for the correlated time series, we made use of the block averaging method.[4] Data series are broken into equal-sized contiguous blocks, and the standard error of the mean in each block is calculated and plotted as a function of increasing block lengths. Then, we obtain the standard error of the mean of the time series by averaging values extrapolated to block lengths of infinite size.

## 1.3. Potential of Mean Force: Adaptive Biased Force (ABF)

The Lennard-Jones free energy profiles (Figure S1) were estimated by means of the Adaptive Biasing Force (ABF) method.[5-9] The mean force, $\langle F_\xi \rangle_\xi$, exerced along the transition coordinate, ξ, can be related to the free energy first derivative according to Eq. 4:

$$\frac{dG(\xi)}{d\xi} = \langle \frac{\delta U(x)}{\delta \xi} - \frac{1}{\beta}\frac{\delta ln|J|}{\delta \xi} \rangle = - \langle F_\xi \rangle_\xi \quad (4)$$

where $|J|$ is the Jacobian determinant to transform the Cartesian coordinates to generalized coordinates, and $\langle F_\xi \rangle_\xi$ denotes the cumulative mean of the instantaneous force, $F_\xi$, in bins δξ wide. Within the ABF formalism, the external force applied, that opposes the free energy barriers to be overcome along the transition coordinate, ξ, can be given by:

$$F^{ABF} = - \langle F_\xi \rangle_\xi \nabla_x \xi \quad (5)$$

where the $F^{ABF}$ in Eq. 5 stands for the biasing force that, ideally, reaches zero in its convergence. The statistical uncertainty of the mean force in each δξ bin was estimated from the standard error of the mean [9] according to the following relation

$$Err\left[\langle F_\xi \rangle_i\right] = \sqrt{\frac{\langle \Delta F_\xi^2 \rangle_i}{m_i}} \quad (6)$$

where $\langle \Delta F_\xi^2 \rangle_i$ represents the variance of the instantaneous force collected in bin $i$, and $m_i$ denotes the independent samples. $m_i$ is calculated as the product between the total number of MD simulation steps, $n_i$, and the MD simulation time step, $\Delta t$, divided by the autocorrelation time, $\tau_i$, of the instantaneous force. In practice, $\tau_i$ was estimated by a non-linear fitting through an exponential function in the form $e^{\frac{-t}{\tau}}$. To calculate the statistical error for points in different windows, we used the Bienaymé formula[10] given as:

$$Err\left[\Delta G_{a \to b}\right] = \delta\xi \left( \sum_{i=i_a}^{i_b} \frac{\tau_i}{n_i \Delta t} \langle \Delta F_\xi^2 \rangle_i \right)^{\frac{1}{2}} \quad (7)$$

## 2. Free energy calculations

### 2.1. ABF simulation details

The PMF profiles for transferring an *uncharged* DOX molecule across the lipid bilayer as a function of the reaction coordinate *z*, defined as the z-component of the distance between the C.O.M. of the *uncharged* solute and the C.O.M. in either a pure PSM or a PSM/cholesterol lipid bilayer, are reported in Figure S1. To obtain the LJ free energy profiles, the reaction coordinate was discretized into five consecutive windows, each 5 Å wide. To keep the solute in the desired membrane depths, the C.O.M. of the uncharged DOX molecule was harmonically restrained in both the window's edges and the xy-plane perpendicular to the bilayer normal with force constant of 50 kcal/mol/Å².

### 2.2. TI simulation details

The positive net charge of DOX$^+$ was counterbalanced by a negative uniformly-distributed background charge of the same magnitude spread over all water molecules to yield the system's electroneutrality. This approach mitigates simulation artifacts that might arise from charged solutes restrained in highly inhomogeneous systems[11], such as lipid bilayers. DOX$^+$ was slowly dragged from the bulk-water phase at different depths along the membrane normal where it was harmonically restrained with force constant of 50 kcal/mol/Å². After insertion and restrain of the solute at the desired bilayer region, the system was pre-equilibrated 10 ns long with the solute in each of these chosen regions in order to provide the simulation starting-point for the subsequent electrostatic free energy estimations. The electrostatic free energies of charging DOX$^+$ at different regions in the lipid bilayer systems were estimated from 20 ns long MD simulations at the end-point states ($\lambda = 0$ and $\lambda = 1$) and at the intermediate state half-way the alchemical electrostatic-coupling path ($\lambda = 1/2$) (Figures S2-S12). We justify the choice of these particular thermodynamic states relying on previous works that investigated and validated optimal ways of calculating electrostatic free energy differences from numerical MD simulations with a reduced number of thermodynamic states being evaluated.[3,12-15] This approach has been proven computationally cheap and accurate for estimating the electrostatic component in free energy differences, such as those of interest in many biological processes where the changes in the electrostatic interactions are the mainly driving force.

**3. MD simulation analysis**

**3.1. Radial Distribution Function**

To obtain the radial distribution profiles, we made use of the *Radial Distribution Function* plugin implemented in VMD (version 1.9.3).[16] The number of particles within a given distance *r* from a reference selection was computed and then normalized by the bin volume and the number density of the bulk-water phase.

**3.2. Hydrogen Bonding**

The H-bond analysis was carried out using the *Hydrogen Bonds* plugin available in VMD. We counted as a H-bond formation when the following geometrical criteria were satisfied: (1) distance between the hydrogen donor (H-donor) and the hydrogen acceptor (H-acceptor) heavy atoms is less than 3.5 Å; (2) the angle formed between the H-donor and H-acceptor, with the hydrogen atom position as vertex, is less than 30°.

## 3.3. Electron Density Profile

To estimate the one-dimension electron density profiles, we collected time-averaged histograms of atomic positions in equally-sized bins of thickness 1.0 Å along the z-direction to the membrane normal. The number density values at each bin were normalized with the total number of electrons assigned to each atom type in the system, heuristically deduced from their respective atomic numbers. The estimated profiles assume that the electron density is located at the atomic center, a widely used approximation that reasonably reproduces experimental observables estimated from x-ray and neutron diffraction measurements.[17]

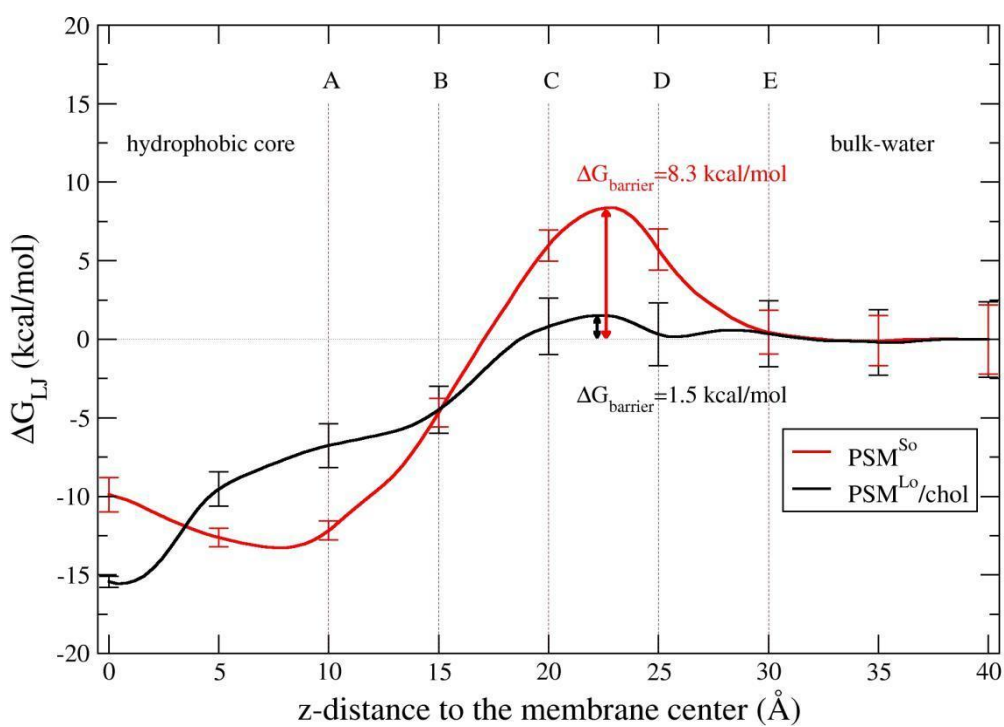

Figure S1. Lennard-Jones free energy profiles for transfering DOX$^+$ from bulk-water phase towards the pure PSM$^{So}$ (red curve) and binary PSM$^{Lo}$/cholesterol (black curve) bilayers.

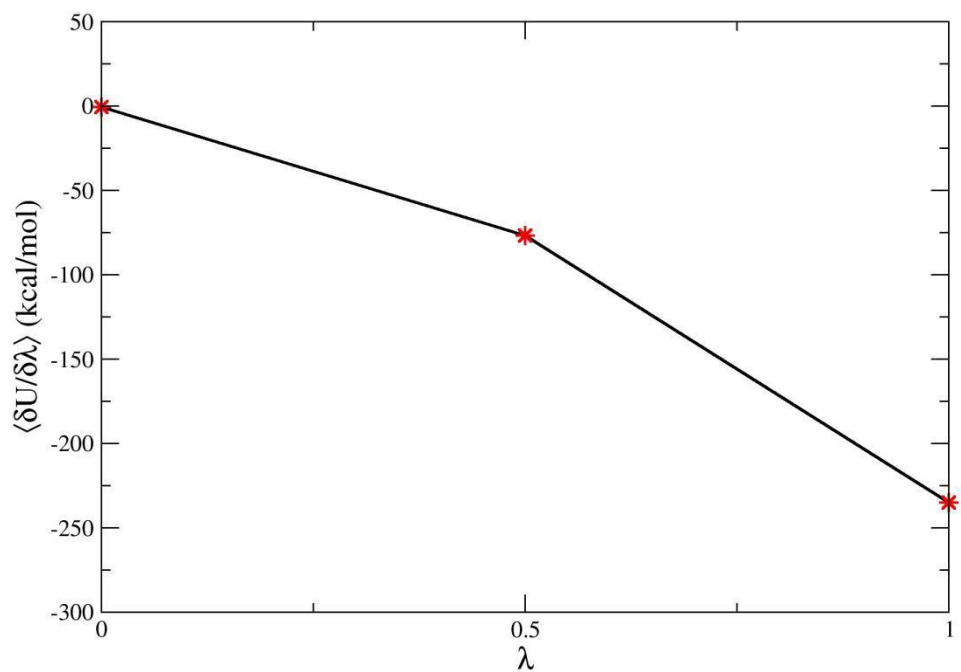

Figure S2. Derivative of the electrostatic potential energy as a function of the coupling parameter lambda for the charging process of $DOX^+$ in the bulk water.

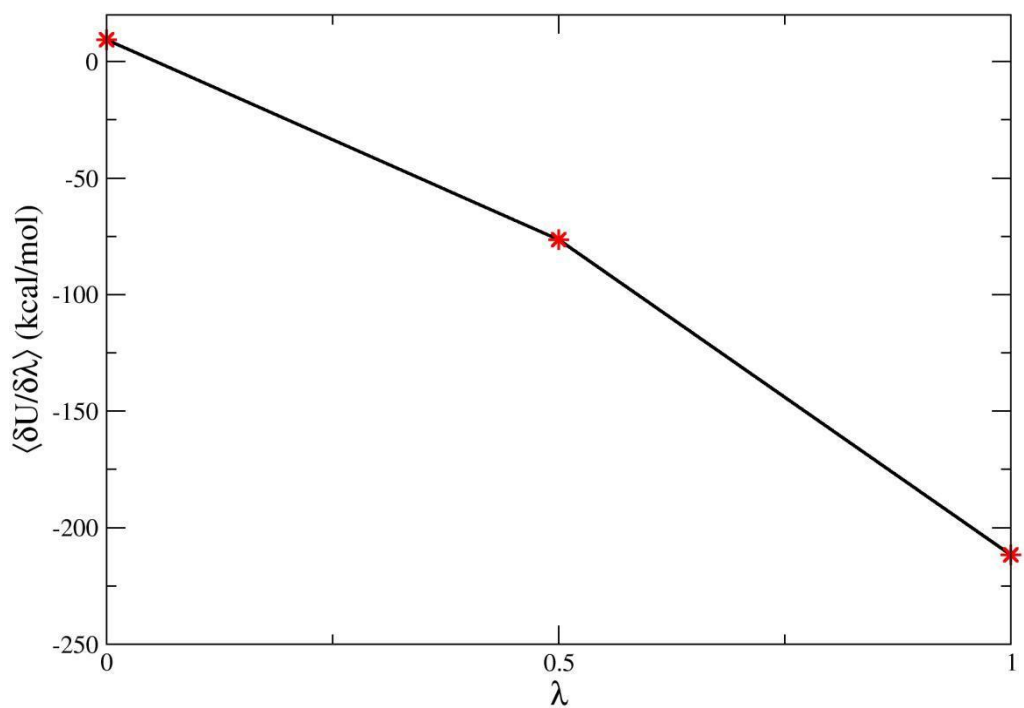

Figure S3. Derivative of the electrostatic potential energy as a function of the coupling parameter lambda for the charging process of DOX$^+$ in the fully hydrated PSM bilayer (Region A).

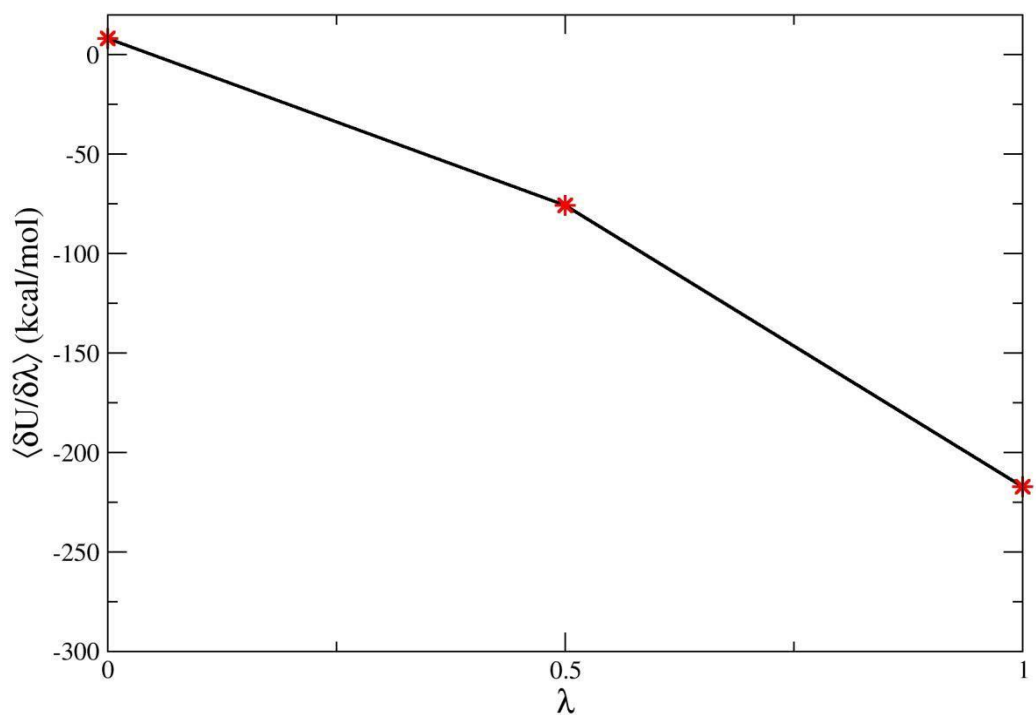

Figure S4. Derivative of the electrostatic potential energy as a function of the coupling parameter lambda for the charging process of DOX$^+$ in the fully hydrated PSM bilayer (Region B).

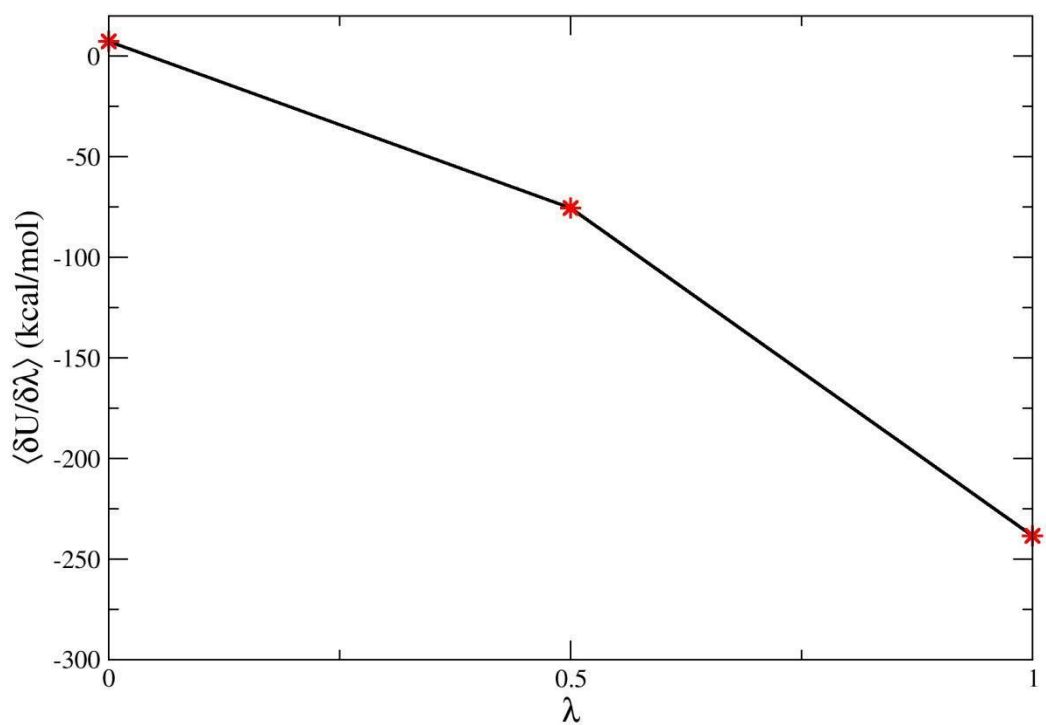

Figure S5. Derivative of the electrostatic potential energy as a function of the coupling parameter lambda for the charging process of DOX$^+$ in the fully hydrated PSM bilayer (Region C).

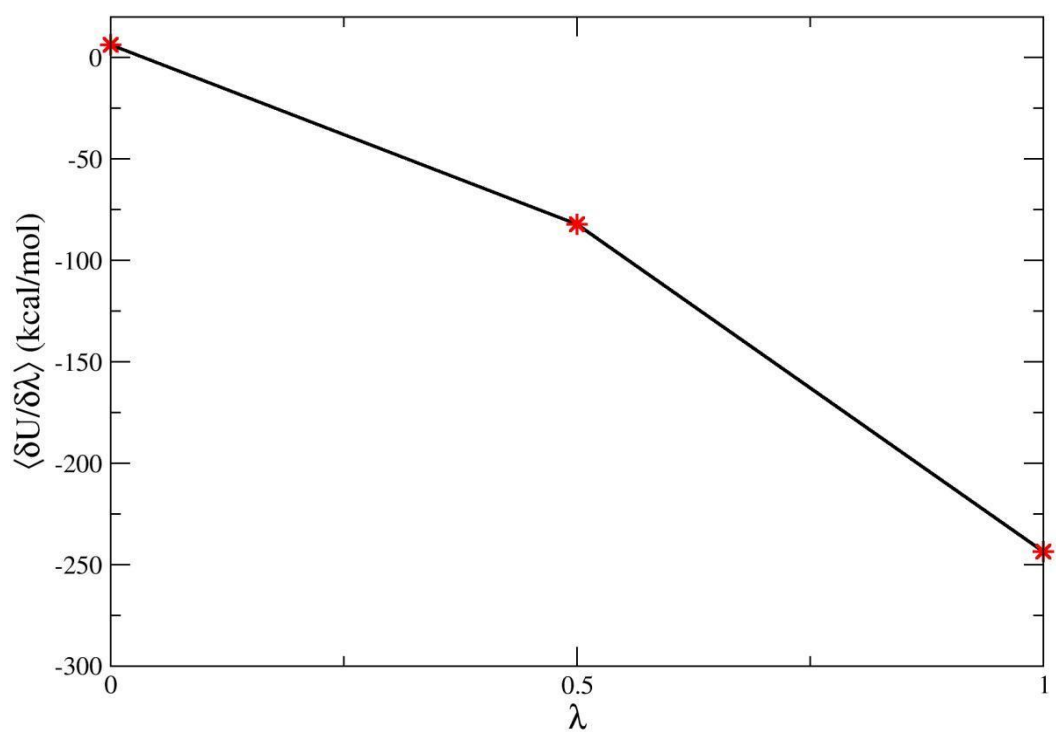

Figure S6. Derivative of the electrostatic potential energy as a function of the coupling parameter lambda for the charging process of DOX$^+$ in the fully hydrated PSM bilayer (Region D).

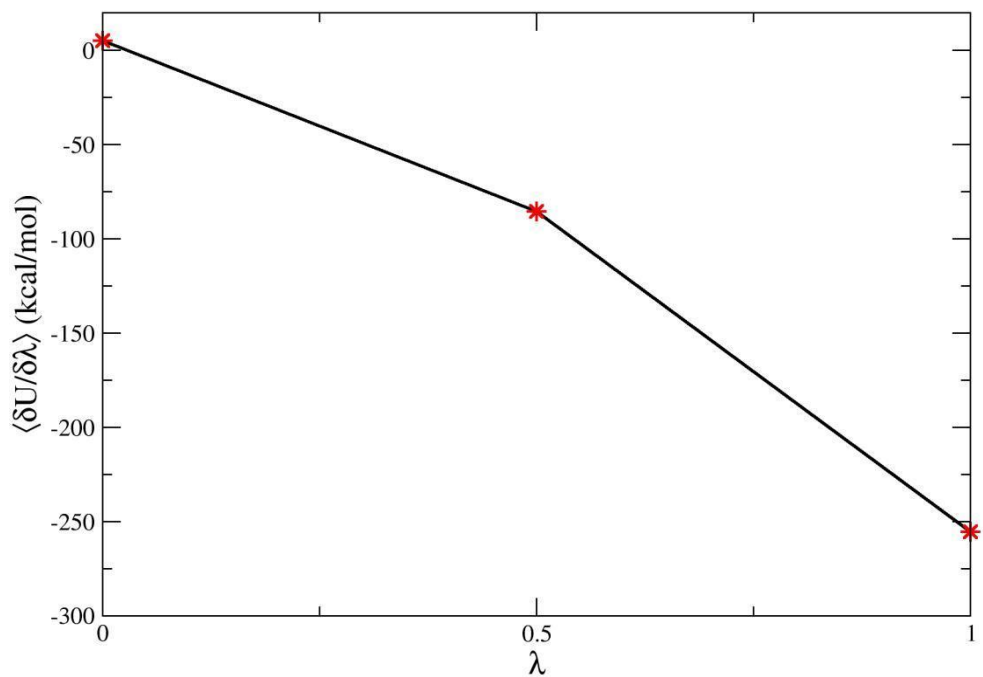

Figure S7. Derivative of the electrostatic potential energy as a function of the coupling parameter lambda for the charging process of DOX$^+$ in the fully hydrated PSM bilayer (Region E).

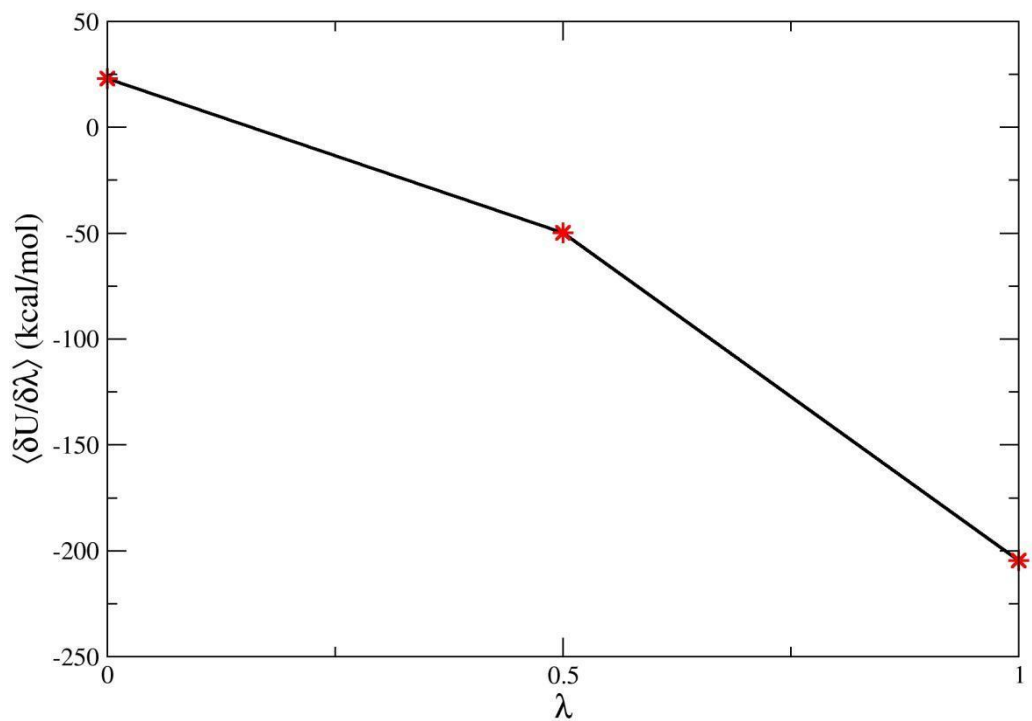

Figure S8. Derivative of the electrostatic potential energy as a function of the coupling parameter lambda for the charging process of DOX$^+$ in the fully hydrated PSM/cholesterol bilayer (Region A).

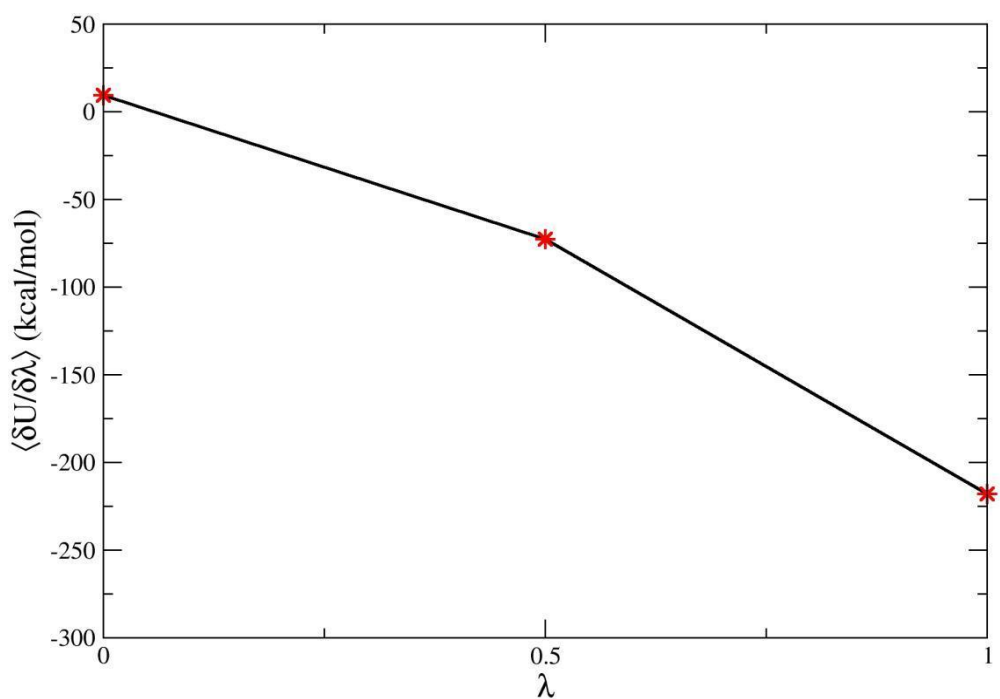

Figure S9. Derivative of the electrostatic potential energy as a function of the coupling parameter lambda for the charging process of DOX$^+$ in the fully hydrated PSM/cholesterol bilayer (Region B).

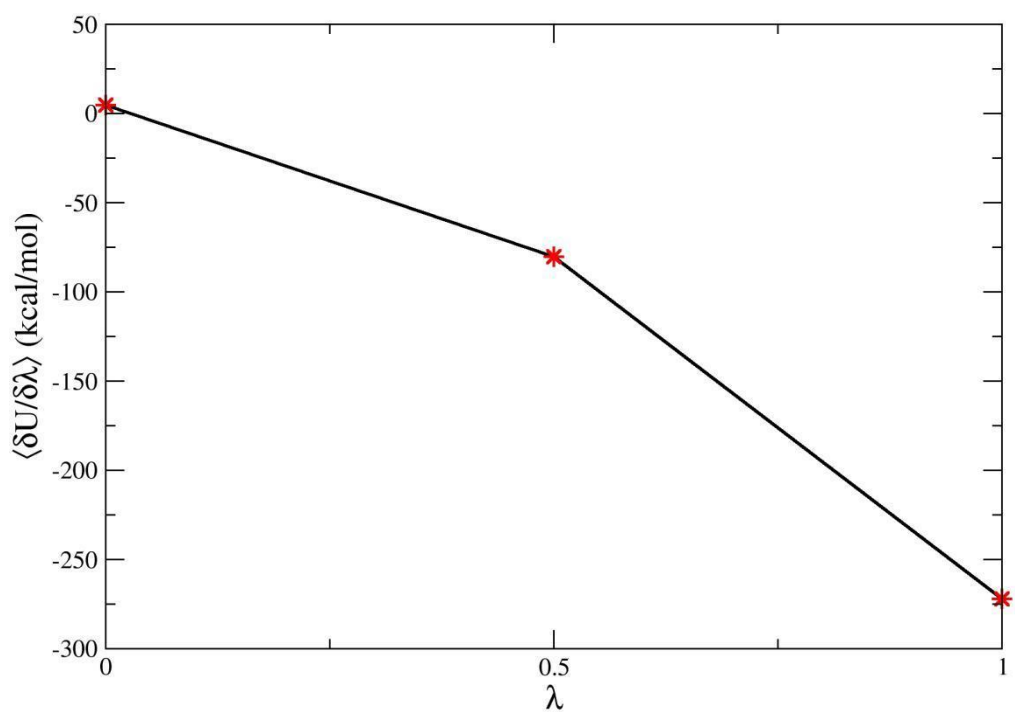

Figure S10. Derivative of the electrostatic potential energy as a function of the coupling parameter lambda for the charging process of DOX$^+$ in the fully hydrated PSM/cholesterol bilayer (Region C).

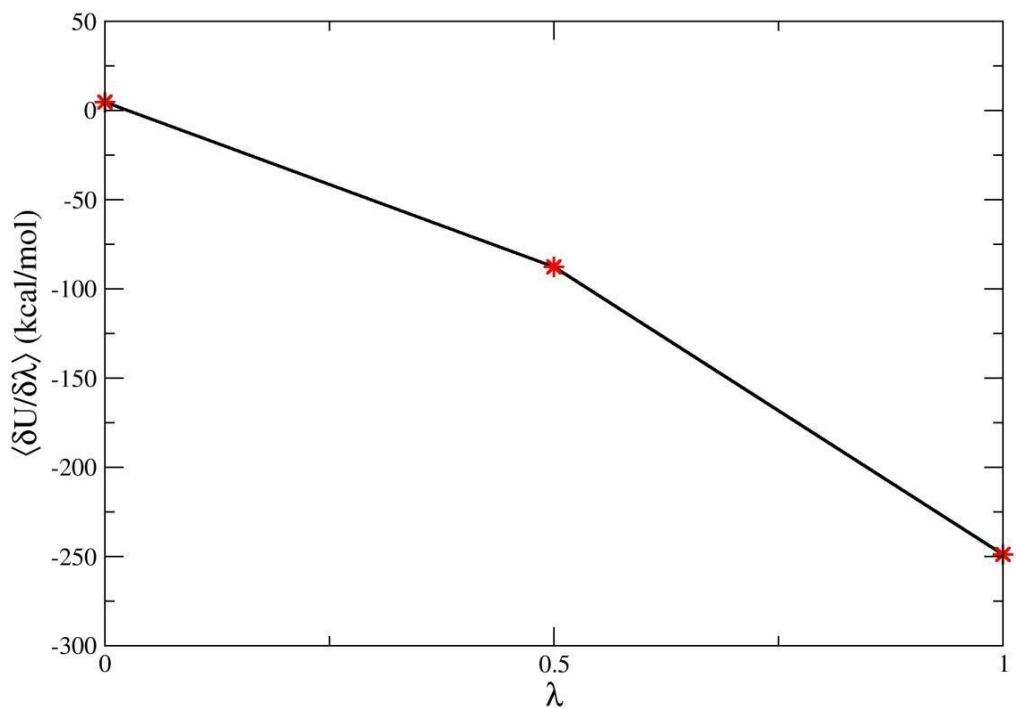

Figure S11. Derivative of the electrostatic potential energy as a function of the coupling parameter lambda for the charging process of DOX$^+$ in the fully hydrated PSM/cholesterol bilayer (Region D).

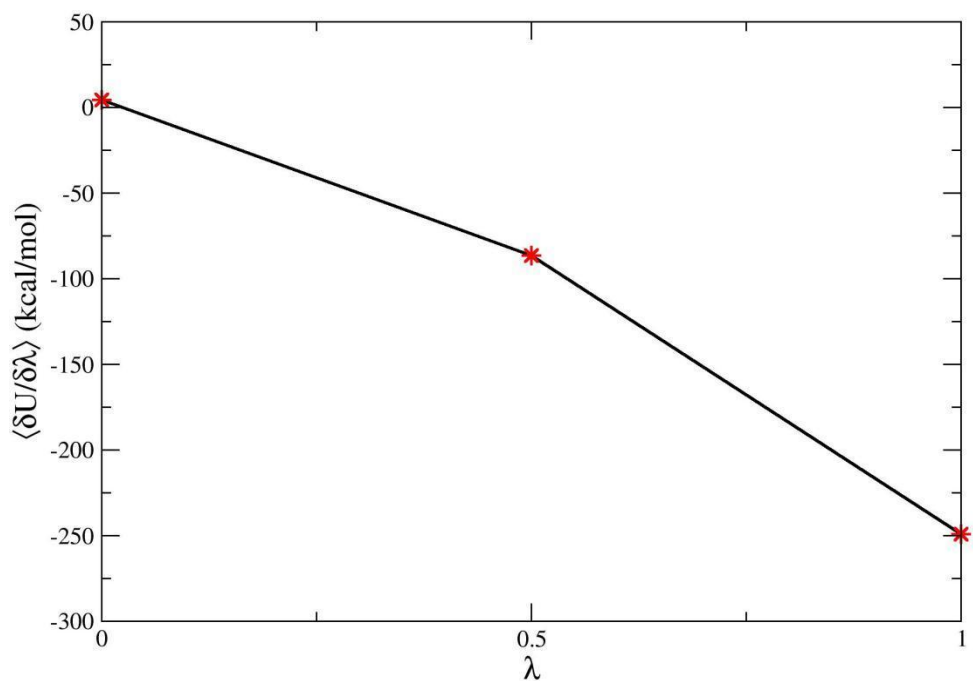

Figure S12. Derivative of the electrostatic potential energy as a function of the coupling parameter lambda for the charging process of $DOX^+$ in the fully hydrated PSM/cholesterol bilayer (Region E).

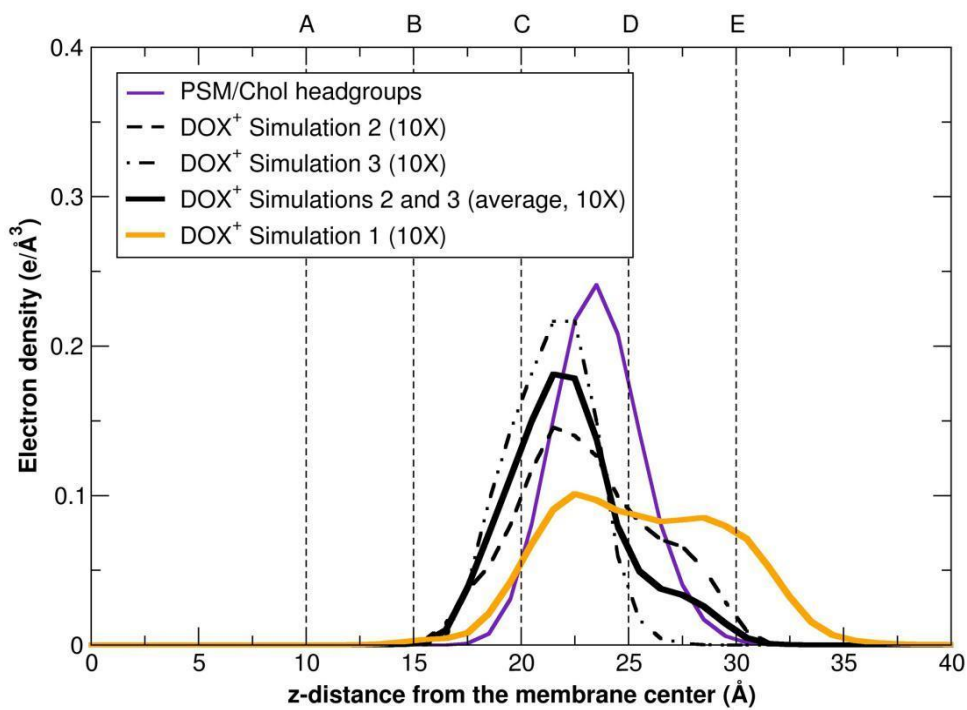

Figure S13. Electron density profile of $DOX^+$ and the $PSM^{Lo}$/cholesterol headgroups over the last 100 ns of production phase from three independent MD simulations 500 ns long. EDP of $DOX^+$ from Simulation 1 (orange line, presented in the manuscript), Simulation 2 (dashed line, first independent MD simulation), Simulation 3 (dash-dotted line, second independent MD simulation)

Figure S13 shows the EDP profiles for a single DOX molecule averaged out of the two independent MD simulations (referred as Simulation 2 and Simulation 3). Figure S13 also includes the EDP profile for DOX as presented in the main text (referred as Simulation 1). We observe that the EDP profile averaged over the last 100 ns of the two independent MD trajectories 500 ns long (Simulations 2 and 3) resembles the one obtained from Simulation 1. These results confirm that the DOX location finds at least two minima at the lipid headgroups region in line with the free energy predictions in Table 1b (main text). However, no convergence of the EDP profiles can be assumed within the time scale sampled in this work.

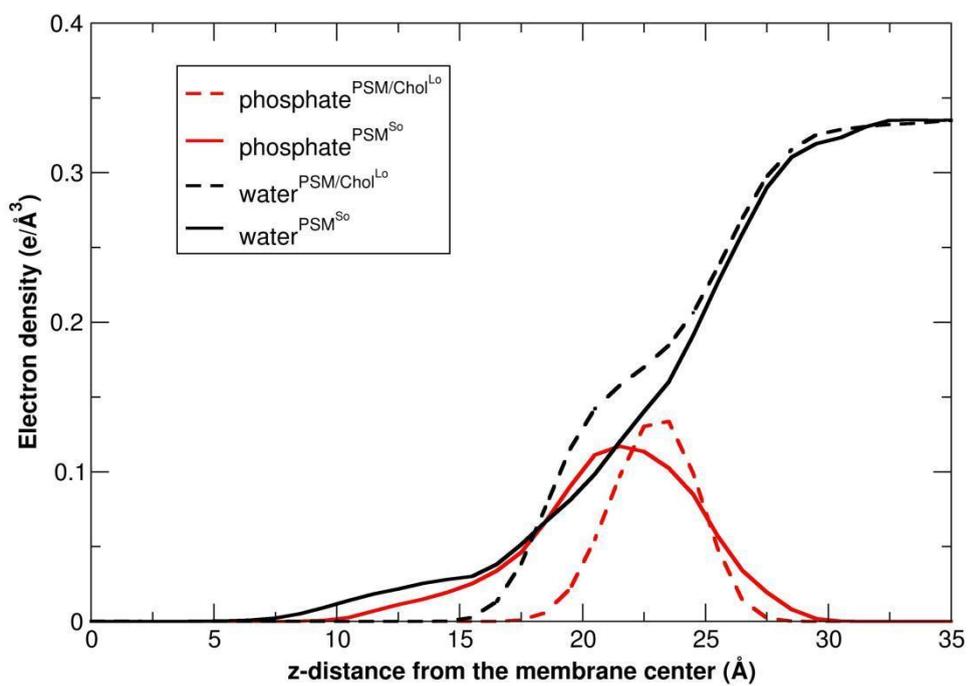

Figure S14. Electron density profile of water (black lines) and phosphate groups (red lines) for the pure PSM$^{So}$ (full lines) and PSM$^{Lo}$/cholesterol (dashed lines) bilayers over the last 100 ns of MD production phase.

Table S1. DOX$^+$ partial atomic charges obtained according to the CHARMM FF protocol. The chemical structure of DOX$^+$ with the atom types assignments is reported, as well.

| Atom type | Charge (e) | Atom type | Charge (e) |
|---|---|---|---|
| O1 | -0.851000 | C24 | -0.116000 |
| O2 | -0.162000 | C25 | -0.115000 |
| O3 | -0.839000 | C26 | -0.112000 |
| O4 | -0.579000 | C27 | -0.100000 |
| O5 | -0.531000 | H1 | 0.420000 |
| O6 | -0.531000 | H2 | 0.377000 |
| O7 | -0.308000 | H3 | 0.420000 |
| O8 | -0.588000 | H4 | 0.420000 |
| O9 | -0.459000 | H5 | 0.377000 |
| O10 | -0.459000 | H6 | 0.404000 |
| O11 | -0.391000 | H7 | 0.404000 |
| N | -0.351000 | H8 | 0.404000 |
| C1 | 0.077000 | H9 | 0.090000 |
| C2 | 0.289000 | H10 | 0.090000 |
| C3 | -0.185000 | H11 | 0.090000 |
| C4 | -0.177000 | H12 | 0.090000 |
| C5 | 0.003000 | H13 | 0.090000 |
| C6 | -0.003000 | H14 | 0.090000 |
| C7 | 0.298000 | H15 | 0.090000 |
| C8 | -0.208000 | H16 | 0.090000 |
| C9 | 0.244000 | H17 | 0.090000 |
| C10 | 0.164000 | H18 | 0.090000 |
| C11 | 0.126000 | H19 | 0.090000 |
| C12 | 0.119000 | H20 | 0.090000 |
| C13 | 0.377000 | H21 | 0.090000 |
| C14 | 0.104000 | H22 | 0.090000 |
| C15 | 0.026000 | H23 | 0.090000 |
| C16 | 0.033000 | H24 | 0.090000 |
| C17 | -0.282000 | H25 | 0.115000 |
| C18 | 0.116000 | H26 | 0.115000 |
| C19 | 0.412000 | H27 | 0.115000 |
| C20 | 0.410000 | H28 | 0.090000 |
| C21 | 0.024000 | H29 | 0.090000 |
| C22 | 0.028000 | H30 | 0.090000 |
| C23 | 0.217000 | | |

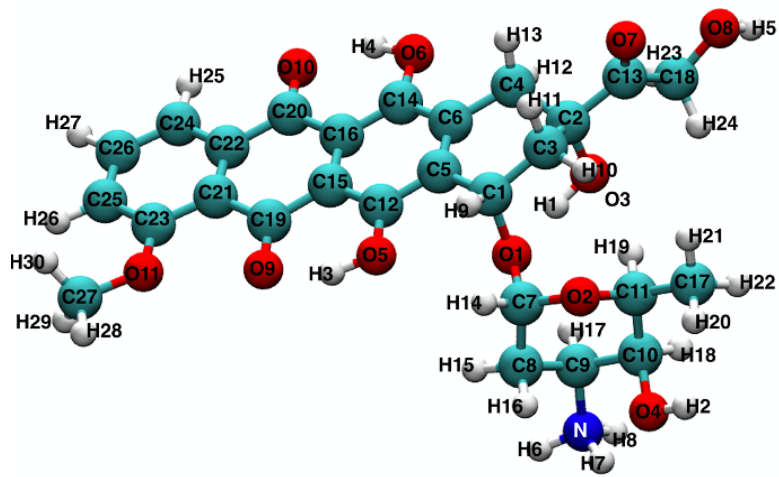

Table S2. DOX$^+$/lipids average non-bonding interaction energies evaluated along the last 100 ns of the molecular dynamics simulations involving the DPPC- and PSM-based bilayers.

| System | DOX$^+$/lipids non-bonding interaction energies (10$^2$ kcal/mol) 400-500 ns | | |
|---|---|---|---|
| | Electrostatic | Van der Waals | Total |
| DPPC$^{So}$/DOX$^+$ | -1.6 (± 0.5) | -0.06 (± 0.05) | -1.6 (± 0.5) |
| DPPC$^{Ld}$/DOX$^+$ | -1.5 (± 0.5) | -0.39 (± 0.06) | -1.9 (± 0.5) |
| DPPC$^{Lo}$/Chol/DOX$^+$ | -0.7 (± 0.5) | -0.26 (± 0.08) | -1.0 (± 0.5) |
| PSM$^{So}$/DOX$^+$ | -1.4 (± 0.6) | -0.03 (± 0.04) | -1.4 (± 0.6) |
| PSM$^{Ld}$/DOX$^+$ | -0.9 (± 0.5) | -0.4 (± 0.1) | -1.3 (± 0.6) |
| PSM$^{Lo}$/Chol/DOX$^+$ | -0.5 (± 0.7) | -0.2 (± 0.1) | -0.7 (± 0.6) |